\documentstyle[12pt,epsf,psfig]{article}

\def\beq{\begin{equation}}
\def\eeq{\end{equation}}
\def\beqa{\begin{eqnarray}}
\def\eeqa{\end{eqnarray}}

\def\a {{\rm f}}

\begin{document}

\begin{flushright}
EDINBURGH 97/22\\
ITP-SB-97-78
\end{flushright}

\begin{center}{\bf\Large\sc Threshold Resummation for}
\vglue 0.25cm
{\bf\Large\sc  Dijet Cross Sections}
\vglue 1.2cm
\begin{sc}
Nikolaos Kidonakis\\
\vglue 0.5cm
\end{sc}
{\it Department of Physics and Astronomy\\
University of Edinburgh,
Edinburgh EH9 3JZ, Scotland, UK}\\
\vskip 1.0cm
\begin{sc}
Gianluca Oderda and George Sterman\\
\vglue 0.5cm
\end{sc}
{\it Institute for Theoretical Physics \\
SUNY at Stony Brook,
Stony Brook, NY 11794-3840, USA}\\
\end{center}
\vglue 1cm
\begin{abstract}
We construct  dijet differential cross sections 
at large momentum transfer, in which threshold 
logarithms have been summed to all orders in perturbation theory.
This extends previous work on heavy quark production,
by treating collinear singularities associated with
hard, massless partons in the final state.
The resummed corrections enable us to
define, in the sense of factorization, the
underlying color exchange mechanism.  The influence of
color exchange on the resummed cross section is contained
in the eigenvalues and eigenvectors of an 
anomalous dimension matrix, which describes the
factorization of coherent soft gluons from the hard scattering.
The precise formulas depend on the partonic scattering angles
and energies, as well as on the  method used to  define
the jets in the final state.  For   
cone dijets at fixed invariant mass, we find leading 
logarithmic corrections that,
like those in the Drell-Yan process, are
positive, and which grow with increasing dijet invariant
mass.  Other choices of dijet cross section can give, however,
qualitatively different behavior, even at leading logarithm.
\end{abstract}
\vfill

\newpage
\section{Introduction}

Factorized cross sections in perturbative QCD 
separate universal parton densities from 
process-specific factors that describe the perturbative hard scattering
of the partons \cite{CSSrv}.  These factors,
or hard-scattering functions, may include singular distributions,
defined by
their integrals with smooth functions, such as the parton densities
relevant to the process.   To be
specific, for the production of a  system of mass $M$,
through the collision of partons of invariant mass
squared $\hat{s}$, the hard-scattering function
will contain, at $n$th order, terms 
as singular as $(\alpha_s^n/n!)[\ln^{2n-1}(1-z)/(1-z)]_+$,
with $z\equiv M^2/\hat{s}$.  The limit $z\rightarrow 1$ is
the edge of partonic phase space, or
partonic threshold, at which no
energy remains for QCD radiation.
These singular contributions often increase the cross
section.  The classic case is the
production of Drell-Yan pairs, with $M=Q$ the invariant
mass of the electroweak vector boson produced by the
annihilation of a quark pair.  
The possibility 
of controlling such singular distributions to all
orders in perturbation theory was recognized early \cite{DYearly}, and
the full analysis, including nonleading
logarithms was carried out  \cite{St87,CT1}.  The formalism   
is closely related to the even earlier resummation of
logarithms in the transverse momenta of the pairs \cite{CoSo81}.  Finally,
it was noted that the resummation in logarithms
of $1-z$ can be performed, not only for the fully inclusive
Drell-Yan cross section $d\sigma/dQ^2$, but also for
the cross section at fixed rapidity, $d\sigma/dQ^2 dy$ \cite{LS}.
We will have occasion to recall and employ the explicit
forms of these resummations below.

In this paper, we shall derive threshold resummation for
several physically-relevant dijet cross sections at large
momentum transfer.    
We will find both similarities
and differences compared to Drell-Yan and related
cross sections, due to the color content of partons
emerging from the hard scattering.  One 
difference is that in jet production the hard scattering 
is itself a QCD subprocess, and involves color exchange.
Another is that the kinematics at threshold depend upon
the method used to define the event.  We shall see that both 
of these features of jet production influence higher-order
corrections in an essential way.  The broad outlines
of our reasoning, however, follow the derivation of
resummation from factorization, as discussed in Ref.\ \cite{CLS}.

When a hard process is mediated by QCD,
rather than color-singlet electroweak annihilation,
the situation becomes, not surprisingly, more complex. Nevertheless,
it is not difficult to recognize that the leading threshold
logarithms associated with the initial state
are the same for QCD processes as for
Drell-Yan at each order in perturbation theory.
This observation was made the basis for the
first estimates of the effects of resummed perturbation theory
in heavy quark production \cite{QQbarLSvN,QQbarBC,QQbarCMNT}.  The extension to nonleading logarithmic
corrections has been carried out recently for
heavy quarks \cite{KS1,KS,KV}.  At this level, intriguing effects begin to
show up, in which the color structure of the underlying
hard scattering influences  the pattern of soft radiation
near threshold.  These results are related to the long-standing
program of developing observable consequences of QCD coherence,
in which properties of hadronic final states reflect color
flow in partonic subprocesses \cite{CoCoh}.

For dijet (as opposed to heavy quark) production,
threshold singularities also differ from the Drell-Yan
case because of the presence of the final-state jets, of momentum
$p_i$, $i=1,2$.  
When the cross section is evaluated at fixed dijet total invariant
mass, $(p_1+p_2)^2$,
these differences turn out to be at the level of next-to-leading
logarithm in the singular distributions.  
As we shall show, however, when the cross section is evaluated
at fixed $(p_1 \cdot p_2)$, new
leading-logarithmic contributions arise, which are negative.
Depending on the partonic subprocess, 
these new contributions may
even overcome the enhancements from
initial state interactions, and produce
an overall suppression of the cross section, relative to 
lowest order.  

The resummation of singular distributions at partonic threshold
is important because corrections very close to
the edge of phase space may be numerically large,
and taken at leading logarithm often (although not always) grow, rather
than decrease, with the order \cite{KS1}.  The resummed cross
section also requires integrals over the argument of the
running coupling \cite{St87,CT1}, indicating enhanced sensitivity to
soft gluon effects, and suggesting the necessity of
 incorporating nonperturbative corrections.  
In addition, the resummation of threshold singularities
in QCD hard scattering requires a general approach to the
interplay of color exchange at short distances with the
pattern of gluon radiation into the final state \cite{CoCoh}.  
We shall see that beyond leading logarithms
it is in general not possible to separate initial from
final state emission, and that, in fact, the color
structure of the hard scattering directly influences
the flow of energy into the final state, not only
at low order, but to all orders in perturbation theory.

As in Ref.\ \cite{KS}, we shall make
strong use of the factorization properties of 
gauge theory cross sections at high energy.  
The basic complication in these arguments
arises from the divergences of massless field theories in Minkowski
space, in which the separation of long- and short-distance
dynamics must be carried out relative to the light-like
directions intrinsic to the problem.  The
principles and methods underlying the arguments below have been
developed, for instance, in Refs.\ \cite{CSSrv,GS78,gsbook}.
We shall try, however, to make the discussion relatively
self-contained, and to explain our use of technical results
as they arise.    

We begin in Sec.\ 2  by identifying the cross sections, based
on cone algorithms for jets, which we shall 
use as illustrative examples in
this paper.  In the
following two sections, we
go on to discuss the factorization properties 
of these cross sections (Sec.\ 3), and then to give interpretations
for the various functions that appear in the 
factorized cross section in terms of field-theoretic
matrix elements of nonlocal operators (Sec.\ 4).  The pattern of
these arguments is close to that used previously for
resummation in heavy quark cross sections \cite{KS},
but now modified to treat the extra collinear divergences
associated with the final-state jets.  
In Sec.\ 5, we organize the singular behavior 
of each of the component functions of the factorization formula.
As in the case
of heavy quarks, we identify an anomalous dimension matrix,
which controls color-sensitive gluon radiation into the
final state\footnote{In a forthcoming paper we will compute this matrix at one-loop 
order and discuss its diagonalization for the basic parton processes,
including gluon-gluon scattering.}.  Combining these results,
we give the resummed expressions for dijet cross sections
in moment space.  We observe that the form of 
our resummed cross section, Eq.\ (\ref{sigNHSfinal}), does
not depend on the details of the jet-identification algorithm,
and applies to any infrared-finite dijet cross section at 
large momentum transfer.
We conclude with a short summary, which looks towards future work.

\section{Cone-based Dijet Cross Sections}

\subsection{Definitions}

\begin{figure}
\centerline{\epsffile{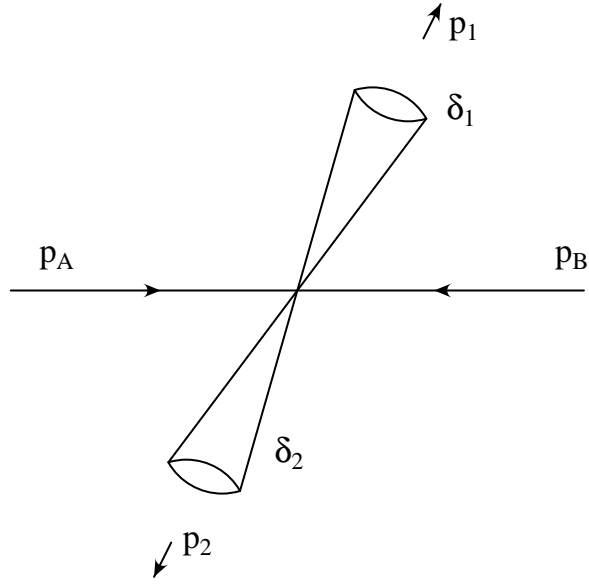}}
\caption{Configuration of cones in the partonic center of mass
frame at partonic threshold.}
\label{cones}
\end{figure}

In this section, we introduce dijet
cross sections, describing  the inclusive hadronic  production
 of a pair of jets, 
\beq
h_A(p_A)+h_B(p_B){\longrightarrow}J_1(p_1)+J_2(p_2)+X(k)\, ,
\label{process}
\eeq
 at fixed rapidity
 interval,
\beq
{\Delta}y={1\over 2}\; \ln\left( {p_1^+\; p_2^- \over p_1^-\; p_2^+}\right)\, ,
\label{deltay}
\eeq 
and total rapidity, 
\beq
y_{JJ}= {1\over 2}\; \ln\left( {p_1^++p_2^+ \over p_1^-+p_2^-}\right)\, .
\label{yjj}
\eeq
We consider the situation represented in 
Fig.\ \ref{cones}.  We have two final-state jets, identified by 
cones of opening angles $\delta_1$ and $\delta_2$. 
The four-vector jet momenta $p_i$ are defined as
the total momenta flowing into the cones. 
In the limit of vanishing $\delta_i$, the ${p_i}^{\mu}$
approach light-like momenta.

To define the dijet cross sections, we must specify a large
invariant, $M_{JJ}$, which is held fixed. A natural choice is
the dijet invariant mass,
\beq
M^2_{JJ}=(p_1+p_2)^2\, ,
\label{jetmass}
\eeq
but other specifications are possible. We shall see, in fact,
that the nature of the resummed cross section depends critically 
on this choice. We shall illustrate this point with the alternate
definition
\beq
M^2_{JJ}=2p_1 \cdot p_2 \, ,
\label{jetmassalt}
\eeq
the scalar product of the  two jet momenta.  In either case, large
$M_{JJ}$ at fixed $\Delta y$ implies a large momentum transfer
in the partonic subprocess.

As we integrate over allowable final states,
the actual total momentum flowing into the cones
may differ from a light-like vector by corrections
proportional to the jet opening angle. 
To simplify our discussion, we shall assume that the cones are
small enough so that 
all contributions proportional to
$\delta_i\ll 1$ may be neglected,
but at the same time large enough so that
\beq
\alpha_s(Q)\ln\left[{1\over\delta_i}\right] \ll 1\, ,
\label{deltainequal}
\eeq
where $Q$ is any of the hard scales of the 
cross section, typically the momentum transfer \cite{GS78,gsbook}.

The introduction of cones
removes all the final-state  collinear singularities 
from the partonic cross section, which
is then infrared safe, once the initial-state collinear singularities 
have been factored into universal parton
distribution functions.

We assume that the jets are identified by an algorithm
that samples phase space for sets of particles flowing
into cones of size $\delta_i$. In this search, the jet direction,
and hence the center of the cone, is defined by the total momentum
of the corresponding set of particles.  The details of the jet
identification algorithm are otherwise not important to our
arguments.  Nevertheless, it may be helpful to have
a definite procedure in mind.

To be specific, we define our cone jet algorithms as follows.
Consider a final state $|X \rangle = | \left \{ q_i \right \}_X \rangle$,
consisting of the set $X \equiv \left \{  q_i \right \}_X $ of particles
with momenta ${q_i}^{\mu}$. For any subset of $X$, $x \subset X$, we define
first a unit vector,
\beq
{\hat{n}}_x \equiv \frac{\sum_{j \in x} {\vec{q}}_j}
{ \left | \sum_{j \in x}  {\vec{q}}_j \right |}\, .
\label{eq:versjet}
\eeq 
Given ${\hat{n}}_x$, we define a new set, $x'({\hat{n}}_x,\delta)$,
consisting of those particles flowing into a cone of half-angle $\delta$,
centered about ${\hat{n}}_x$,
\beq
 x' \equiv \left \{ q_k \: : \: \frac{{\vec{q}}_k \cdot {\hat{n}}_x}
{\left| {\vec{q}}_k \right|} \geq \cos \delta \right \}. 
\label{xprimedef}
\eeq
In these terms, a {\it jet}
 is defined to be any set of particles, $x \subset X$,
for which 
\beq
x=x',
\eeq
{\it i.e.}, for which the particles within the cone defined by the set
constitute the set exactly.

Of course, with this definition, a general final state may have many jets.
To define our dijet cross sections, we take ${p_1}^{\mu}$ and  ${p_2}^{\mu}$ to
be the pair of jets of highest energy in the sample, consistent with some
restrictions on their directions, implicit in our choices of ranges for 
$y_{JJ}$ and $\Delta y$ above.
These algorithms are insensitive to emission of zero-momentum lines
and/or rearrangements of momenta among collinear-moving particles. We may therefore expect
them to be infrared safe \cite{GS78,GS79}. 
Of course, as noted in \cite{GS78}, the cancellation of virtual and real
corrections fails when the final state reaches the edge of the region of phase space
that defines a cone-jet cross section.  (For a recent discussion of such
momentum configurations in event shapes, see \cite{CataniWebber}.)
This happens, for example, when a 
finite-energy line 
$ q_l  \in x$, reaches the boundary of the cone about
${\hat{n}}_x$, {\it i.\ e.}, when 
\beq
 \frac{{\vec{q}}_l \cdot {\hat{n}}_x}
{ \left | {\vec{q}}_l \right |} = \cos \delta.
\label{incone}
\eeq
Such singularities, however, are restricted to a 
lower-dimensional surface in phase space, 
and thus do not spoil the finiteness of 
the inclusive jet cross section \cite{GS78}.

\subsection{Factorization}

The standard factorized form of a hard dijet cross section is   
\beqa
\frac{d\sigma_{h_Ah_B{\rightarrow}J_1J_2}(S,\delta_1,\delta_2)}
{dM^2_{JJ}dy_{JJ}d{\Delta}y}&=&
{1\over S^2}\, 
\sum_{f_A,f_B=q,\overline{q},g} 
\int dx_A dx_B\; \phi_{f_A/h_A}(x_A,\mu^2)
\phi_{f_B/h_B}(x_B,\mu^2) \nonumber \\
&\ &\ \ \times H_{f_Af_B}\left( {p_i\cdot p_j \over \mu^2},
\alpha_s(\mu^2),\delta_1,\delta_2\right)\, ,
\label{stdfact}
\eeqa
where  the explicit factor of $1/S^2$ is introduced to make
$H_{f_Af_B}$ dimensionless.
The incoming partons carry fractions $x_A$ and $x_B$ of the momenta 
of the incoming hadrons $h_A$ and $h_B$, respectively.
These integrals are weighted by the 
nonperturbative but experimentally measurable 
parton distribution functions (densities), $\phi_{f/h}$.
 The factorization scale $\mu$ separates the long-distance 
physics described by the parton distributions
 from the short-distance hard scattering.
The first argument, $p_i\cdot p_j/\mu^2$,
of the hard-scattering 
function,
$H_{f_Af_B}$, represents all
large invariants formed from the momenta of the incoming partons and
final-state jets, $p_i,p_j=x_Ap_A,x_Bp_B,p_1,p_2$, $i\ne j$.

The perturbative factors $H_{f_Af_B}$ are smooth functions only 
away from the edges of partonic phase space.  The 
threshold for the partonic subprocess is conveniently
parameterized by  the variable $z$,
\beq
z=\frac{M^2_{JJ}}{x_Ax_B{S}}\equiv \frac{M^2_{JJ}}{\hat{s}}\, ,
\label{eq:zeta}
\eeq
where $S=(p_A+p_B)^2$ is the overall center of mass energy squared,
and $\hat{s}=x_Ax_BS$ is the corresponding quantity for the 
subprocess.  
At $z=1$ (partonic threshold) there is  just enough partonic energy to produce the
observed final state, with no additional radiation.
In general, $H$ includes distributions with respect to $1-z$, with
singularities at
$n$th order in $\alpha_s$ of the type
\beq
\left[ {\ln^{m}(1-z) \over 1-z} \right ]_+, \hspace{10mm} m\le 2n-1\, ,
\label{leadingsing}
\eeq
defined, as usual, by their integrals with any smooth functions ${\cal F}(z)$,
\beqa
\int_y^1 dz \left[ {\ln^{m}(1-z) \over 1-z} \right ]_+\; {\cal F}(z)
&=&
\int_y^1 dz \left[ {\ln^{m}(1-z) \over 1-z} \right ] \left[{\cal F}(z)-{\cal F}(1)\right]\nonumber\\
&\ & \quad - {\cal F}(1)\; \int_0^y dz \left[ {\ln^{m}(1-z) \over 1-z} \right ]\, .
\eeqa
All distributions of this sort have been resummed for Drell-Yan
and more recently for heavy quark
pair production cross sections at leading and nonleading
logarithms \cite{St87,CT1,KS}.
In the following, we extend this analysis to dijet cross sections. 

We can simplify Eq.\ (\ref{stdfact}), 
using the observation of \cite{LS}, which applies 
to QCD as well as electroweak cross sections.  
According to Ref.\ \cite{LS}, in computing 
the leading power ($1/(1-z)$) 
of $H$ for  $z\rightarrow 1$, we may treat the total 
rapidity $y_{JJ}$ of the produced
heavy system (the dijets) as
a constant, equal to its value at threshold, 
$y_{\rm thresh}=(1/2)\ln(x_A/x_B)$,
taking $p_A^\mu=p_A^+\delta_{\mu +}$, $p_B^\mu=p_B^-\delta_{\mu -}$.
In this leading-power approximation, which includes all
logarithmic corrections of the form of (\ref{leadingsing}),
both leading and nonleading, we may write  
\beqa
&&\frac{d\sigma_{h_Ah_B{\rightarrow}J_1J_2}(S,\delta_1,\delta_2)}
{dM^2_{JJ}dy_{JJ}d{\Delta}y}=
{1\over S^2}\, 
\sum_{f_A,f_B=q,\overline{q},g}\int_{\tau}^1dz
\int dx_A dx_B\; \phi_{f_A/h_A}(x_A,\mu^2)
\nonumber \\
&&\hspace{20mm}\times\phi_{f_B/h_B}(x_B,\mu^2)\; \delta
\left(z-\frac{M^2_{JJ}}
{\hat{s}}\right)\delta\left(y_{JJ}-\frac{1}{2}\ln\frac{x_A}{x_B}\right)\nonumber \\
&&\hspace{20mm}\times
\sum_{f_1,f_2}
\hat{\sigma}_{f_Af_B\rightarrow f_1f_2}\left(1-z,\frac{M_{JJ}}{\mu},{\Delta}y,\alpha_s(\mu^2),\delta_1,\delta_2\right)\, .
\label{eq:facto}
\eeqa
In this form, we have introduced $z$, defined in Eq.\ (\ref{eq:zeta}),
as an integration variable, measuring the 
fraction of partonic energy squared going into
the observed final state, in our case the jet pair.
Its lower limit is
\beq
z_{\rm min}\equiv\tau=\frac{M^2_{JJ}}{S}\; ,
\label{taudef}
\eeq
with $M_{JJ}$ defined as in Eq.\ (\ref{jetmass}) or (\ref{jetmassalt}).
With either choice (as noted above)
$z_{\rm max}=1$ corresponds to partonic threshold. 
The simplified, dimensionless, hard-scattering function $\hat{\sigma}$,
now depends only on  $1-z$, $\Delta y$, the ratio of $M_{JJ}$ to $\mu$,
the coupling and the cone angles.  Up to corrections of order $\delta_i$,
each jet evolves from one of two partons, $f_1$ and $f_2$, emerging from
the hard scattering, $f_A+f_B\rightarrow f_1+f_2$, as indicated.  

To compute the hard-scattering function perturbatively,
we turn to (infrared-regularized) parton-parton scattering, which obeys
the same factorization properties as  Eq.\ (\ref{eq:facto}).
The leading power as $z\rightarrow 1$ comes
entirely from flavor diagonal distributions $\phi_{f/f}(x,\mu^2)$ \cite{St87}.
Similarly, at leading power we may sum incoherently over the flavors
of the partons $f_1,f_2$ that fragment into the final state jets.
Thus, for incoming flavors $f_A,f_B$, the partonic hard scattering functions
may be written as a sum over partonic subprocesses
$f_A+f_B \rightarrow f_1+f_2$, as above,
\beqa
&&\frac{d\sigma_{f_Af_B{\rightarrow}J_1J_2}(S,\delta_1,\delta_2)}
{dM^2_{JJ}dy_{JJ}d{\Delta}y}=
{1\over S^2}\, 
\int_{\tau}^1 dz
\int dx_A dx_B\; \phi_{f_A/f_A}(x_A,\mu^2)
\nonumber \\
&&\hspace{20mm}\times\phi_{f_B/f_B}(x_B,\mu^2)\; \delta
\left(z-\frac{M^2_{JJ}}
{\hat{s}}\right)\delta\left(y_{JJ}-\frac{1}{2}
\ln\frac{x_A}{x_B}\right)\nonumber \\
&&\hspace{20mm}\times \sum_{f_1,f_2=q,\overline{q},g}
\hat{\sigma}_{f_Af_B\rightarrow f_1f_2}
\left(1-z,\frac{M_{JJ}}{\mu},{\Delta}y,\alpha_s(\mu^2),
\delta_1,\delta_2\right)\, ,
\label{eq:factoparto}
\eeqa
with the {\em same} hard-scattering functions 
$\hat{\sigma}_{f_Af_B\rightarrow f_1f_2}$ as above.
To calculate $\hat{\sigma}$ to any order of perturbation theory,
we construct the partonic cross section to that order, and
factorize initial state
 collinear divergences into the 
light-cone  distribution functions $\phi_{f/f}$, expanded in $\alpha_s$.
The remainders give the perturbative expansion for the infrared-safe hard
 scattering function \cite{CSSrv}.

Singular distributions of the sort (\ref{leadingsing}) are 
now conveniently organized by 
taking a Mellin transform of the rapidity-integrated partonic cross section 
(\ref{eq:factoparto}) with respect to $\tau$, Eq.\ (\ref{taudef}),
\beqa
\int_0^1 d\tau\; \tau^{N-1}\; \int dy_{JJ}\;
S^2\; \frac{d\sigma_{f_Af_B{\rightarrow}J_1J_2}(S,\delta_1,\delta_2)}
{dM^2_{JJ}dy_{JJ}d{\Delta}y}&\ &\nonumber\\
 &\ & \hspace{-40mm} = \sum_{\a}{\tilde \phi}_{f_A/f_A}(N+1,\mu^2,\epsilon)\; 
{\tilde \phi}_{f_B/f_B}(N+1,\mu^2,\epsilon) \nonumber \\
&\ & \hspace{-35mm}
\times {{\tilde{\sigma}}}_{\a}(N,M_{JJ}/\mu,\alpha_s(\mu^2),\delta_1,\delta_2)\, ,
\nonumber \\
\label{eq:mellin}
\eeqa
with $\tilde{\sigma}(N)=\int_0^1dz\; z^{N-1}\hat{\sigma}(z)$, and 
$\tilde{\phi}(N+1)=\int_0^1dx\; x^{N}\phi(x)$.  To reduce clutter in the
notation, we have denoted the set of $2\rightarrow 2$ partonic reactions  $f_A+f_B\rightarrow f_1+f_2$, 
collectively by $\a$.
Divergent distributions in $1-z$ produce powers of $\ln(N)$, according to
\beq
\int_0^1 dz\; z^{N-1}\left[{\ln^m(1-z)\over 1-z}\right]_+
={-1\over m+1}\ln^{m+1}{1\over N} +{\cal O}\left(\ln^{m-1}N\right)\, .
\label{logmoment}
\eeq
In the following, we shall resum logarithms of $N$, from which the singular
distributions of $\hat{\sigma}$ may be reconstructed by inverting the
transform.  In the limit of large $N$, neglecting terms that decay as $1/N$,
we may replace $N+1$ by $N$ in the arguments of the distribution
functions $\phi$.

\section{Refactorization for Leading Regions}

We are now ready to construct a 
new, ``refactorized", expression for the partonic
cross section, which generates all singular distributions of
the form of Eq.\ (\ref{leadingsing}) at partonic threshold.
It will include, besides functions for the jet pair, a function that describes 
the soft, but still perturbative, radiation outside the cones, and which
responds to the color flow in the hard scattering.  The
basic steps in the factorization process are quite similar
to those already undertaken for Drell-Yan and heavy quark
cross sections \cite{St87,KS}, but the kinematics of jet production
is sufficiently subtle to warrant a description of how the
general treatment must be modified to be applicable in this case.

\subsection{Leading regions and phase space}

We consider the purely partonic reaction
in which flavors $f_A,f_B$, of momenta $p_A$ and
$p_B$ collide to produce the observed jets.
The first step in writing a factorized form for this partonic
cross section at threshold is to identify those
regions of momentum space that contribute to the
cross section at the (singular) leading power of
$1/(1-\tau)$ with $\tau=M_{JJ}^2/S=M_{JJ}^2/(p_A+p_B)^2$.  

\begin{figure}
\centerline{\epsffile{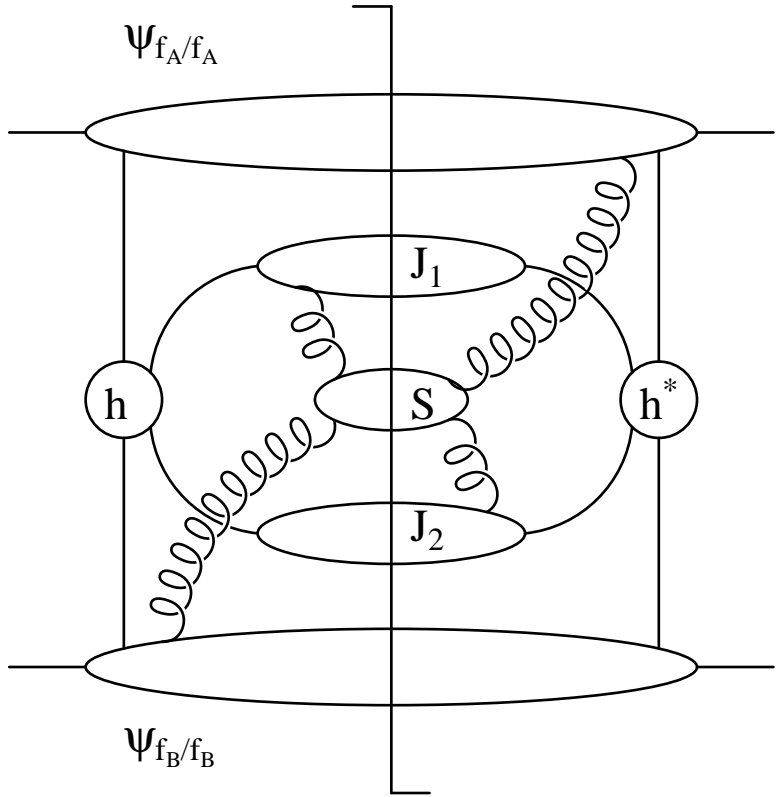}}
\caption{Reduced diagram that represents the generic leading
region for the dijet cross section.}
\label{leadingregions}
\end{figure}

For a two-jet cross section these {\em leading regions} are
illustrated in Fig.\ \ref{leadingregions}, 
which represents the cross section
in terms of a cut diagram \cite{CSSrv,GS78,gsbook}.
Leading power contributions arise from momentum
configurations in which on-shell lines of finite
energy fall into one of  four ``jet" subdiagrams.
Two of these, labelled $\psi_{f_A/f_A}$ and $\psi_{f_B/f_B}$ in the
figure, represent the perturbative evolution of
the incoming partons,  and the motion of their fragments
into the final state.  Logarithms arise from regions in
momentum space where internal lines of the diagrams 
approach the mass shell at the subspaces illustrated
by the figure.  In Fig.\ \ref{leadingregions} and below, we shall
work in a general axial (or temporal), $n\cdot A=0,\, n^2\ne 0$, gauge.  This
simplifies the analysis somewhat, by insuring that
collinear logarithms occur only internally to the 
jets \cite{CSSrv,GS78,gsbook}.  The soft subdiagram, $S$, generates only a single
infrared logarithm per loop in this class of gauges.
We emphasize, however, that our final results will be
gauge invariant, and that it is possible
to reexpress all of our arguments in covariant gauges.

Returning to Fig.\ \ref{leadingregions}, the $\psi$'s
give rise to a pair of active partons, whose
scattering initiates the jet event.
The two ``short-distance functions" $h$ and $h^*$, 
to which the active partons connect,
contain the effects of off-shell partons at the
hard scatterings in the amplitude and its complex
conjugate. 
The two remaining jets, $J_1$ and $J_2$,
represent the fragmentation of the partons
emerging from the hard scatterings
into the observed jets.  
Finally, the function $S$ represents
the soft radiation, coupling incoming and outgoing
hard partons.  

The leading regions pictured in Fig.\ \ref{leadingregions} are
identified by means of the analyticity
and power counting techniques described in Refs.\ \cite{CSSrv,GS78,gsbook}.  Compared to
the leading regions of the Drell-Yan process, which they generalize,
they differ by the presence of the observed final-state jets, to which
soft radiation may couple, and which require a
sum over color structures for the hard scatterings.
To anticipate, the
total partonic cross section may be factorized
into functions $\psi_i$, $J_i$, $h$ and $S$
corresponding to these quanta.  We shall have
more to say below about how and why this may
be done, and how to define the singular
functions explicitly as
vacuum expectation values of nonlocal,
composite operators in QCD.

We shall follow the procedure developed for 
the Drell-Yan process in Ref.\ \cite{St87} by constructing the 
jet and soft functions in such a way that they
absorb all singular dependence on $1-\tau$.  This
can be done in a straightforward manner, by
matching the phase space of partons included
in the $\psi$'s, $J$'s and $S$ with the phase
space of the underlying process.  
It is easiest to carry out this analysis in
the center of mass (c.m.) frame of the incoming partons.

Consider first exact threshold for the two-jet
process.
At partonic threshold in the c.m.\ frame, the jets $J_1$ and $J_2$ carry 
equal and opposite-moving momenta, $p_1$ and $p_2$,
\beq
p_1^\mu+p_2^\mu= (M_{JJ},\vec{0})= p_A^\mu+p_B^\mu\, .
\label{p1p2}
\eeq
Now suppose we emit soft radiation of total momentum
$k$ outside the jet cones.  The total squared invariant mass 
necessary for this radiation in the partonic process
is 
\beq
S \equiv (p_A+p_B)^2=(p_1+p_2+k)^2=(p_1+p_2)^2+2M_{JJ}k_0
+{\cal{O}}(M_{JJ}^2(1-\tau)^2)\, ,
\label{thresholdkin}
\eeq
where we have used Eq.\ (\ref{p1p2}).  In terms of
the variable $\tau$, Eq.\ (\ref{taudef}), we then have,
\beqa
1-\tau&=&{2k_0 \over M_{JJ}}+{\cal O}((1-\tau)^2) \hspace{26mm} \left( M^2_{JJ}=(p_1+p_2)^2 \right)\, \nonumber \\
1-\tau&=&\frac{{p_1}^2+{p_2}^2}{ M^2_{JJ}}+{2k_0 \over M_{JJ}}
+{\cal O}((1-\tau)^2) \hspace{10mm} \left( M^2_{JJ}=2p_1 \cdot p_2 \right)\,.
\label{k0to1minusz}
\eeqa
Thus near threshold, when $M^2_{JJ}=(p_1+p_2)^2$, the total invariant mass of the two jets,
the  variable $1-\tau$ is given
by the total c.m.\ energy radiated outside
the cones, up to corrections
suppressed by a power of $1-\tau$, which 
 we have agreed to neglect.
On the other hand, for $ M^2_{JJ}=2p_1 \cdot p_2$, the variable $1-\tau$ receives contributions from the 
jet masses as well. This is a crucial difference, and leads to different behavior near threshold
at the leading logarithmic level. The reason for this difference is that partonic threshold in the
variable $M^2_{JJ}=2p_1 \cdot p_2$ is surprisingly more restrictive than in the variable $M^2_{JJ}=(p_1+p_2)^2$.
In the latter case, since we
sum over all particles and momenta emitted into the cones,
there is already considerable phase space available for
the production of a pair of jets of the specified
invariant mass ($M_{JJ}$), total rapidity and scattering
angles.  The analogs of these thresholds in ${\rm e}^+{\rm e}^-$ annihilation
are jet configurations in which all energy
is emitted into two opposite-moving jet cones, with no
soft radiation in the intervening directions.  The
sum over all such states has infrared, but {\em not}
collinear divergences, as the latter cancel
in the sum over allowed states.  Then, when an
infrared energy resolution $\delta E$ is introduced, and
soft radiation outside the cone is summed over up to 
$\delta E$, we find one logarithm of $\delta E$
per loop \cite{gsbook}.  

For the case $M^2_{JJ}=2p_1 \cdot p_2$, 
Eq.\ (\ref{k0to1minusz}) shows that $\tau=1$ requires vanishing jet invariant mass
in addition to energy flow only into the cones. Many fewer final states 
then contribute at threshold, specifically 
only those final states which consist of massless jets (at the cones' centers,
according to the construction of Sec.\ 2.1).
It is not surprising that large corrections arise as a result. Because 
they reflect a restriction in phase space,
the additional corrections in this case are negative,
corresponding to incomplete cancellation of
collinear singularities. 

The refactorized cross section is pictured in 
Fig.\ \ref{factfig}, where the double lines represent, as
we shall see below, propagators associated with
ordered exponentials of the gluon field in the jet
directions.  
This decoupling of soft gluons from jet subprocesses
is at the basis of proofs of factorization for 
single-particle inclusive cross sections \cite{CoSt80}
and of factorization for inclusive cross sections \cite{factproofs,CSS88}.
We shall discuss the
technical justification for the form of factorization
proposed here
in the next subsection.  
The physical basis of jet/soft factorization, however, is the inability
of soft gluons to resolve the internal substructure of the jets.
The effect of all attachments of soft gluons to the jets
is summarized simply by gauge rotations on the partons that
connect the jets to the hard scatterings \cite{CSSrv,CoSo81}.  The gauge rotations
may be represented by the ordered exponentials of the
gauge fields that we shall encounter below. 

Given the structure in Fig.\ \ref{factfig},
we may write the partonic cross section 
$d\sigma_{f_Af_B\rightarrow J_1J_2}$
in a preliminary factorized form, as the product of functions 
which describe the short- and long-distance
processes in the scattering,
\beqa
\frac{d\sigma_{f_Af_B{\rightarrow}J_1J_2}(S,\delta_1,\delta_2)}
{dM^2_{JJ}d{\Delta}y}
&=&        {1\over S^2}\, 
\sum_{\a}\sum_{IL} 
H_{IL}^{(\a)}\left({M_{JJ}\over\mu},\Delta y,\alpha_s(\mu^2)\right)\; 
\nonumber \\
&\ & \quad
\times
{K}_{LI}^{(\a)} 
\biggl ( {M_{JJ}\over \mu },{(1-\tau)M_{JJ}\over \mu },\Delta y,
\alpha_s(\mu^2),\delta_1,\delta_2 \biggr )  
\nonumber \\ 
&\ & \quad \quad +{\cal O}(1-\tau)\, .  
\label{prelimfact}
\eeqa
Here $H_{IL}$ includes the hard scatterings.
The overall rapidity, $y_{JJ}$, is integrated over (although for $\tau$ near 1,
it is always small
in the c.m.\ system).  At the same time, both the long-
and short-distance factors depend upon the rapidity
difference $\Delta y$, or equivalently, the c.m.\ scattering angle.
The  indices $I,L$ describe possible
 color structures at the hard vertices,
for the scattering  amplitude and for
the complex conjugate amplitude. 
We may further factorize the fully short-distance function
into contributions from the amplitude and its complex conjugate,
\beq
H_{IL}^{(\a)}\left({M_{JJ}\over\mu},\Delta y,\alpha_s(\mu^2)\right) 
=
{h^*}_{L}^{(\a)}\left({M_{JJ}\over\mu},\Delta y,\alpha_s(\mu^2)\right)\;
h_{I}^{(\a)}\left({M_{JJ}\over\mu},\Delta y,\alpha_s(\mu^2)\right)\, .
\nonumber\\ 
\label{uvjetfact}
\eeq
As $\tau\rightarrow 1$, only purely virtual loops can carry momenta
of order $M_{JJ}^2$.  These are the contributions kept in $H_{IL}$.
The color structures $L$ and $I$, of course,
depend implicitly on the flavor content of the hard scattering, denoted by $\a$.  
For example, when the
underlying partonic hard-scattering is $q\bar{q}\rightarrow q\bar{q}$, 
there are only two possible color 
structures, which may be chosen as singlet and octet
in the $s$ or $t$ channel.
Similarly, the color-dependent factors $h$ depend as well
upon the spin of outgoing scattered lines.
The dependence on spin, however, follows
the same pattern as dependence on flavor, and  we  shall not
exhibit it explicitly.  We always assume unpolarized collisions.

It is also useful to factorize the
long-distance part of the cross section, $K_{LI}$,
separating partons that are collinear to the incoming
quarks from those that are in the final-state jets
and those that are soft and ``central" in
rapidity.  Referring to Eq.\ (\ref{k0to1minusz}), the
sum of the weights associated with
the functions is $(1-\tau)$, so that
the convolution is of the form
\beqa
{K}_{LI}^{(\a)}
\biggl ( {M_{JJ}\over \mu },{(1-\tau)M_{JJ}\over \mu },\Delta y,
\alpha_s(\mu^2),\delta_1,\delta_2 \biggr )  
&\ & 
\nonumber \\
&\ & \hspace{-55mm} =
\int_0^1 {dw_A}\; {dw_B}\; dw_1\; dw_2\;
{dw_S}\; \delta(1-\tau-w_A-w_B-w_1-w_2-w_S)\nonumber \\
&\ & \hbox{\hskip -50mm}\times
{\psi}_{f_A/f_A}\left ( w_A,{M_{JJ}\over \mu },\alpha_s(\mu^2),\epsilon \right )
{\psi}_{f_B/f_B}\left (w_B,{M_{JJ}\over \mu },\alpha_s(\mu^2),\epsilon \right )
\nonumber\\
&\ & \hbox{\hskip -50mm} \times 
{S}_{LI}^{(\a)} \biggl ( {w_SM_{JJ}\over \mu },\Delta y,\alpha_s(\mu^2) \biggr )
\nonumber\\
&\ &  \hspace{-50mm} \times\; 
J^{(f_1)}\left(w_1,{M_{JJ}\over\mu},\alpha_s(\mu^2),\delta_1\right)\;
J^{(f_2)}\left(w_2,{M_{JJ}\over\mu},\alpha_s(\mu^2),\delta_2\right)
\nonumber\\
&\ &\quad  +{\cal O}\left(1-\tau\right)\, .
\label{KpsipsiS}
\eeqa
Dependence on the gauge vector $n^\mu$ is present in each
factor, but has been suppressed.
The five functions $\psi_A$, $\psi_B$, $J^{(f_1)}$, $J^{(f_2)}$
and $S_{LI}^{(\a)}$ are all convoluted 
together in terms of
the weights that each function contributes to 
the final state.  
This is possible because the weights identified
in Eq.\ (\ref{k0to1minusz}) above are additive in the particles of
the final state, and because the factorization implies
that there is no interference between final-state particles
associated with the different functions.
Let $k_S^0$ be the energy of particles emitted by
the soft function $S$ outside the cones, and similarly
for the four other functions.  Then the lack of interference
implies that
\beq
k^0=k_A^0+k_B^0+k_1^0+k_2^0+k_S^0\, ,
\label{knaughtsum}
\eeq
with $k^0$ the energy in Eq.\ (\ref{k0to1minusz}).
The complete contributions of each of the functions to the weights are given by
\beqa
w_A={2k_A^0\over M_{JJ}}\quad &\ & w_B={2k_B^0\over M_{JJ}}\nonumber\\
w_1={p_1^2+2M_{JJ}k_1^0\over M_{JJ}^2}\quad &\ & 
w_2={p_2^2+2M_{JJ}k_2^0\over M_{JJ}^2}\nonumber\\
w_S &=&{2k_S^0\over M_{JJ}}
\label{functionws1}
\eeqa
for $M_{JJ}^2=2p_1\cdot p_2$, and by
\beqa
w_A={2k_A^0\over M_{JJ}}\quad &\ & w_B={2k_B^0\over M_{JJ}}\nonumber\\
w_1={2k_1^0\over M_{JJ}}\quad &\ & 
w_2={2k_2^0\over M_{JJ}}\nonumber\\
w_S &=&{2k_S^0\over M_{JJ}}\, ,
\label{functionws2}
\eeqa
for $M_{JJ}^2=(p_1+p_2)^2$. 
Note that the product 
$\psi_A\psi_BJ_1J_2S$ of jet and soft functions behaves as $(w_Aw_Bw_1w_2w_S)^{-1}$ 
when $1-\tau$ vanishes \cite{St87,CLS}, which is how logarithmic
enhancements arise.  We will be able to neglect
contributions to $p_i^2$ except from $J^{(f_i)}$
because we neglect corrections proportional to $\delta_i$.
Finally, the dependence on the jet opening angles is included 
entirely in the jet functions, up to corrections 
proportional to the $\delta_i$.  Although it is possible
to do so, we shall not attempt a resummation in the 
opening angles (recall our assumption, Eq.\ (\ref{deltainequal})).
Before exploring the
consequences of these factorized expressions, let us discuss 
their justification.

\subsection{Relation to factorization for inclusive processes}

The details of the proof of the factorization of soft quanta from a jet
depend on whether the jet in question is ``initial-state", as the $\psi_i$
in Fig.\ \ref{factfig}, or ``final-state", as the $J_i$.  Factorization
is somewhat simpler in the latter case, and was discussed in
Ref.\ \cite{CoSt80} in the context of single-particle inclusive
annihilation, and in Ref.\ \cite{CoSo81} in the context
of transverse momentum distributions.  The arguments are
essentially identical in this case.  The factorization
of soft interactions from initial-state jets is an essential
ingredient in proofs of factorization for inclusive 
processes, and was extensively discussed in Refs.\
\cite{factproofs,CSS88} in this connection.  

In factorization proofs for inclusive cross sections, the goal
is to show that soft interactions cancel, but the
factorization of soft quanta from jets is an important
subsidiary result in this demonstration \cite{CSS88}.  
It is necessary, however, to distinguish the factorization
of soft quanta from partonic jets in Eq.\ (\ref{KpsipsiS}),
in which $z$ is fixed near unity (near threshold), from
their factorization in the inclusive case, where $z$ is freely integrated over.
Let us recall a few
of the relevant arguments of Ref.\ \cite{factproofs,CSS88},
as they apply here.
First, the difference between initial-state and final-state jets lies
in the singularity structures of the jets, in terms of
the momentum components of soft quanta.  For final-state jets,
all poles are in the same half-plane, corresponding
to final-state interactions only.  This simplifies
the analysis considerably.  Indeed, a short argument
based on contour deformation 
allows us to use the eikonal, or ``soft", approximation for
soft gluon momenta within jets, and to replace all the
details of the jets' interactions with soft gluons by
ordered exponentials \cite{CSSrv,CoSo81}.  

For initial-state jets,
however, contour deformation is severely restricted,
on a graph-by-graph basis, as poles from final-state
{\it and} initial-state
interactions conspire to pinch the momentum integration
contours in regions for which the eikonal approximation
fails \cite{CoSt80}.  It is, in general, only after the sum over
final states that final-state interactions cancel,
the pinches disappear, and soft quanta may be factored.
When $z$ is fixed near unity, or when moments are
taken, final states are not all summed over with the
same weight, and we may not, in general, factor
soft quanta from jets in the same manner as in Ref.\ \cite{CSS88},
 without leaving over apparently nonfactoring
remainders.  
In our case, however, such remainders are
simply absorbed into the definition of the soft function
$S_{LI}$, through Eq.\ (\ref{KpsipsiS}). 

The only properties of $S_{LI}$ that we shall need are that 
it is infrared safe and that 
it has at most a single, infrared logarithm per loop
in the limit $w_S\rightarrow 0$.
Its infrared safety follows from the universality of collinear
singularities, which have been absorbed into the $\psi_i$'s.
Its single-logarithmic dependence on  $w_S$ requires further explanation, however.

Double logarithms require both collinear and infrared enhancements.
Collinear logarithms must arise from momentum
configurations in which lines in $S_{LI}$ that are
parallel to the initial-state jets are much more energetic
than soft lines that connect the two jets.
But this requires that the soft lines are much
softer in energy than $w_SM_{JJ}$, since the total,
and hence the jet energy is bounded by this quantity.
We may then apply the reasoning of Ref.\ \cite{CSS88}
to any such momentum configuration to show that such
a region {\em cancels}, until the soft gluons become
energetic enough so that they may no longer be ignored kinematically 
compared to lines that are parallel to the incoming
momenta.  But in this region, we cannot generate collinear logs,
and we expect at most
one logarithm per loop.  We shall review the explicit forms
of the soft and jet functions in the following section.
First, however, we discuss the role of moments.
 
\subsection{Moments of the partonic cross section}

From Eqs.\ (\ref{prelimfact}) and (\ref{KpsipsiS}), we
find for the full partonic cross section,
\beqa
\frac{d\sigma_{f_Af_B{\rightarrow}J_1J_2}(S,\delta_1,\delta_2)}
{dM^2_{JJ}d{\Delta}y}
&=& {1\over S^2}\, \sum_{\a}\sum_{IL} 
H_{IL}^{(\a)}\left({M_{JJ}\over\mu},\Delta y,\alpha_s(\mu^2)\right)
\nonumber \\
&\ & \hspace{-40mm}
\times \int_0^1 {dw_A}\; {dw_B}\;
dw_1\; dw_2\; {dw_S}\; \delta(1-\tau-w_1-w_2-w_A-w_B-w_S)
\nonumber \\
&\ &  \hspace{-40mm}
\times {\psi}_{f_A/f_A}\left ( w_A,{M_{JJ}\over \mu },\alpha_s(\mu^2),\epsilon \right )
\;
{\psi}_{f_B/f_B}\left (w_B,{M_{JJ}\over \mu },\alpha_s(\mu^2),\epsilon \right )  \nonumber\\
&\ & \hspace{-40mm}
\times {S}_{LI}^{(\a)} \biggl ( {w_SM_{JJ}\over \mu },\Delta y,\alpha_s(\mu^2) \biggr )\nonumber\\
&\ & \hspace{-40mm}  \times J^{(f_1)}\left(w_1,{M_{JJ}\over \mu },\alpha_s(\mu^2),\delta_1\right)\; 
J^{(f_2)}\left(w_2,{M_{JJ}\over \mu },\alpha_s(\mu^2),\delta_2\right)\nonumber\\
&\ & \hspace{-30mm} \quad
+{\cal O}(1-\tau)\, , 
\label{eq:facthatsig}
\eeqa
where, as above, $\tau=M_{JJ}^2/S$.
In moment space, this becomes
\beqa
\int_0^1 d\tau\; \tau^{N-1}\; S^2\;
\frac{d\sigma_{f_Af_B{\rightarrow}J_1J_2}(S,\delta_1,\delta_2)}
{dM^2_{JJ}d{\Delta}y} &=&         
\sum_{\a}\sum_{IL} 
H_{IL}^{(\a)}\left({M_{JJ}\over\mu},\Delta y,\alpha_s(\mu^2)\right)\nonumber\\
&\ &\ \hspace{-30mm}\times\; {\tilde\psi}_{f_A/f_A}\left ( N,{M_{JJ}\over \mu },\alpha_s(\mu^2),\epsilon \right )
\;
{\tilde\psi}_{f_B/f_B}\left (N,{M_{JJ}\over \mu },\alpha_s(\mu^2),\epsilon \right )  \nonumber\\
&\ &\ 
\hspace{-30mm}\times\; {\tilde S}_{LI}^{(\a)} \biggl ( {M_{JJ}\over \mu N},\Delta y,\alpha_s(\mu^2) \biggr ) 
\nonumber \\ 
&\ &\ 
\hspace{-30mm}\times\; {\tilde J}^{(f_1)}\left(N,{M_{JJ}\over \mu },\alpha_s(\mu^2),\delta_1\right)\; 
{\tilde J}^{(f_2)}\left(N,{M_{JJ}\over \mu },\alpha_s(\mu^2),\delta_2\right)\nonumber \\
&\ & 
\hspace{-30mm}\quad  +{\cal O}(1/N)\, .  
\label{eq:factimp}
\eeqa
Comparing Eqs.\ (\ref{eq:mellin}) and  (\ref{eq:factimp}) 
we derive for the Mellin transform of the hard scattering function
the ``refactorized" expression, accurate to $O(1/N)$,
\beqa
\tilde{{\sigma}}_{\a}(N)&=&
\left[\frac{ {\tilde{\psi}}_{f_A/f_A}(N,M_{JJ}/\mu,\epsilon)
{\tilde{\psi}}_{f_B/f_B}(N,M_{JJ}/\mu,\epsilon)}
{{\tilde{\phi}}_{f_A/f_A}(N,\mu^2,\epsilon)
{\tilde{\phi}}_{f_B/f_B}(N,\mu^2,\epsilon)}
 \right]\nonumber \\
& &\times\sum_{IL} 
H_{IL}^{(\a)}\left({M_{JJ}\over\mu},\Delta y,\alpha_s(\mu^2)\right)\; 
{\tilde S}_{LI}^{(\a)} \biggl ( {M_{JJ}\over \mu N},\Delta y,\alpha_s(\mu^2) \biggr )
\nonumber \\
&\ &\ \times 
{\tilde J}^{(f_1)}\left(N,{M_{JJ}\over \mu },\alpha_s(\mu^2),\delta_1\right)\; 
{\tilde J}^{(f_2)}\left(N,{M_{JJ}\over \mu },\alpha_s(\mu^2),\delta_2\right)\, .
\label{sigNHS}
\eeqa
The first factor is ``universal" between electroweak and
QCD-induced hard processes, and was computed first with $f_A=q$ for
the Drell-Yan cross section \cite{St87}.  

To organize both the $\mu$- and $N$-dependences of the 
jet and soft functions, we develop
definitions for them in terms of matrix elements
in the next section.  We will then go on to
show how the renormalization of these matrix elements
leads to the resummation of logarithms of $N$.

\section{Soft and Jet Functions}

We now turn to the explicit forms of the jet and soft functions that 
result from factorization, in terms of matrix elements of
composite operators.  It is the renormalization properties
of these operators that will lead to resummation of
threshold singularities, in the next section.

\subsection{Center of mass distribution}

 We begin with the center of mass parton distribution functions $\psi_{f/f}$ \cite{St87}, 
appearing in the ``refactorized" expression, Eq.\ (\ref{eq:facthatsig}),
where they absorb long-distance contributions of the initial-state jets,
while respecting the overall phase space restrictions near partonic threshold. 
The functions $\psi$ differ from standard light-cone parton distributions by being
defined at fixed energy, rather than 
light-like momentum fraction.  They were introduced for 
the inclusive Drell-Yan cross section, and applied as
well to heavy quark production \cite{KS}.  For completeness, we
may define them by analogy to light-cone parton distributions
via the matrix elements,
\beqa
\psi_{q/q}(x,2p_0/\mu,\epsilon)
&=&
{1\over 2\pi 2^{3/2}}
\int_{-\infty}^\infty dy_0\ {\rm e}^{-ixp_0y_0}
\langle q(p)|\bar{q}(y_0,\vec{0})\; {1\over 2}v\cdot \gamma\; q(0)|q(p)\rangle\nonumber \\
\psi_{\bar{q}/\bar{q}}(x,2p_0/\mu,\epsilon)
&=&
{1\over 2\pi 2^{3/2}}
\int_{-\infty}^\infty dy_0\ {\rm e}^{-ixp_0y_0}
\langle \bar{q}(p)|\;
{\rm Tr}\left[\; {1\over 2}v\cdot \gamma\; q(y_0,\vec{0})
\bar{q}(0)\right]\; |\bar{q}(p)\rangle\nonumber \\
\psi_{g/g}(x,2p_0/\mu,\epsilon)
&=&
{1\over 2\pi 2^{3/2}p^+}
\int_{-\infty}^\infty dy_0\ {\rm e}^{-ixp_0y_0}
\langle g(p)|v_\mu F^{\mu\perp}(y_0,\vec{0})\; v_\nu {F^\nu}_\perp(0)|g(p)\rangle\, ,\nonumber\\
\label{psidef}
\eeqa
where the matrix element is evaluated in $n\cdot A=0$ gauge
in the partonic c.m.\ frame (in Ref.\ \cite{St87},
$A_0=0$ gauge was chosen).  The vector $v$ is light-like in
the opposite direction from $p^\mu$, so that for $\vec{p}$ in
the $\pm 3$ direction, $v\cdot\gamma=\gamma^\pm$.  
In the antiquark distribution, the combination $q\bar{q}$ is treated
as a Dirac matrix to define the trace.  Charge conjugation invariance
implies that $\psi_{q/q}=\psi_{\bar{q}/\bar{q}}$.
The  moments of these distributions  may be factorized into 
a product of moments of the light-cone parton distribution $\phi$, defined in any scheme,  times
an infrared safe function \cite{St87}.  The argument $\epsilon$ in $\psi$ 
represents the universal collinear singularities that $\psi$
absorbs in the factorized
cross section, Eq.\ (\ref{eq:facthatsig}).

\subsection{Matrix elements for final-state jets}

As the second step in the construction of the functions into
which the cross section factorizes, we treat the 
final-state jets $J^{(f_i)}$.
In axial gauges, all collinear logarithms are generated by
the imaginary parts of two-point functions.  
In the factorized cross section, Eq.\ (\ref{eq:facthatsig}),
the final-state jets are linked to the other
functions through the weight convolution.
The only direct dependence on the momentum of
final-state particles associated with $J^{(f_i)}$
is in the hard-scattering function $H_{IL}^{(\a)}$,
which depends on the total momenta
emitted into the cones through $\Delta y$ and $M_{JJ}$.   In
$H_{IL}$, the final-state (and initial-state) jet
momenta are approximated by light-like momenta
$p_i^{(0)}{}^\mu=\beta_i^\mu M_{JJ}/\sqrt{2}$, 
with $\beta_i$ a light-like
velocity vector in the direction of the jet.  
The vector $p_i^{(0)}$ is determined by
the three-momentum of 
particles flowing into the jet cone at partonic threshold.
In constructing the jet function, we will therefore
sum over the complete phase space of the particles
associated with $J^{(f_i)}$, subject to
fixed total spatial momentum of particles
within the cone, {\it and} fixed weight $w_i$.

To be specific, let  $\ell_i$ be the
total momentum of the final-state particles of
$J^{(f_i)}$, $p_i$ be the total momentum of
particles of $J^{(f_i)}$ emitted into cone $i$,
and $k_i$ the total momentum of the particles of $J^{(f_i)}$
emitted outside cone $i$, so that $\ell_i=p_i+k_i$.
The sum over phase space includes the integral over
all $k_i$, and over the invariant mass $p_i^2$,
at fixed values of $\vec{p}_i{}^{(0)}$, defined as above.
The remaining phase space is that of two light-like
vectors $p_i^{(0)}$, one for each of the two jets.
These vectors are fixed in our differential cross section
for $M_{JJ}$ and $\Delta y$, Eq.\ (\ref{eq:facthatsig}).
Overall factors associated with this two-particle
phase space are implicitly absorbed into the
definition of the $H_{IL}$.

Consider, for example, an outgoing quark.  Near threshold, the
jet function is the cut quark two-point Green function, summed over final
states at fixed $w_i$ and $\vec{p}_i{}^{(0)}$.  
This two-point function depends, in general, 
on $\vec{p}_i{}^{(0)}$, and on
the gauge vector, $n^\mu$.  
It may thus be defined as
\beqa
J^{(f)}_{\beta\alpha,ba}\left(\vec{p}_i{}^{(0)},{w_i},
M_{JJ},\mu,\alpha_s(\mu^2),\delta_i\right) 
&=& \sum_{\xi}\; 2|\vec{p}_i{}^{(0)}|\; 
(2\pi)^3\delta^3\left({\vec{p}_i{}^{(0)}}-\vec{p}_\xi\right)  \nonumber\\
&\ & \hspace{-30mm} \times\; \delta\left(w_i-w(\xi,\delta_i)\right)\;    
\langle 0|\; f_{\beta,b}(0)\;
|\xi \rangle \langle \xi|\;
 \bar{f}_{\alpha,a}(0)\; |0\rangle\, , \nonumber\\
\label{eq:outjet}
\eeqa
where $f_{\beta,b}$ is the field of flavor
$f$, with Dirac and color
indices $\beta$ and $b$, respectively.  As usual, we
suppress dependence on $n^\mu$.
The factor $2|\vec{p}_i{}^{(0)}|\; 
(2\pi)^3\delta^3\left({\vec{p}_i{}^{(0)}}-\vec{p}_\xi\right)$ in 
Eq.\ (\ref{eq:outjet}) matches the
normalization of $p_i{}^{(0)}$ phase space, and fixes  $\vec{p}_\xi$, the spatial
momentum of all particles in state $\xi$ that flow
into the cone.  
The sum is over all states $\xi$ with the specified jet momentum, 
consistent with a contribution $w_i$
to the overall weight.  

In our examples, the weight
is given by either the first or second line
in Eq.\ (\ref{k0to1minusz}).  With either definition,
as $w_i$ vanishes, the momenta of all particles outside
the cone are forced to vanish.
In the case $M_{JJ}^2=2p_1\cdot p_2$, for example,
final-state particles from $J_i^{(f_i)}$ contribute through
the term $p_i^2/M_{JJ}^2$
if they are within the cone, and through $2k_i^0/M_{JJ}$ if they are outside
the cone, as in Eq.\ (\ref{functionws1}).  
The function $w(\xi,\delta_i)=p_i^2/M_{JJ}^2+2k_i^0/M_{JJ}$ 
is simply 
the sum of these two contributions, where $k_i^0$ is the total
energy of final state particles in $\xi$ that flow outside the
cone $\delta_i$.  
In our approximation, we may neglect corrections to $J^{(f_i)}$
that are proportional to  $\delta_i$,
the jet invariant mass, $p_i^2$, and $w_i$.  

Near partonic threshold, the invariant mass of jet $i$ is
given by
\beq
p_i^2 \simeq 2p_i^{(0)}\cdot p_i=\sqrt{2}M_{JJ}\beta_i\cdot p_i\, ,
\label{psubisquare}
\eeq
where again $\beta_i$ is a light-like vector in
the jet direction.  
In a frame where $\vec{\beta}$ is in the $+3$ direction,
 $\beta_i\cdot p_i$ is the minus 
(opposite-moving) light-cone 
component of the total momentum of the particles emitted into the cone.
(Recall that by the jet definition of Sec.\ 2.1,
$\vec{p}_{i,T}\equiv 0$; see Eq.\ (\ref{eq:versjet}).)

Another consequence of our approximations is that for a quark jet
the only Dirac matrix structure that we need to retain is
proportional to $\gamma\cdot p^{(0)}$,
\beqa
J^{(f_i)}_{\beta\alpha,ba}
\left(\vec{p}_i{}^{(0)},w_i,M_{JJ},\mu,\alpha_s(\mu^2),\delta_i\right) &\ &\nonumber\\
&\ & \hspace{-40mm} =\left(\gamma\cdot p_i^{(0)}\right)_{\beta\alpha}\delta_{ba}\,
J^{(f_i)}\left(w_i,{M_{JJ}\over\mu},\alpha_s(\mu^2),\delta_i\right)\, ,
\label{sigmamatrix}
\eeqa
where the function $J^{(f_i)}$ without indices is
a scalar distribution in $w_i$.  
This is the function in the factorized cross section
Eq.\ (\ref{eq:facthatsig}).
Corrections to (\ref{sigmamatrix})
are suppressed by a power of $w_i$.
Similar considerations apply to gluon jets.

 \subsection{The eikonal cross section}

Having identified matrix elements
that describe initial-state and final-state jets,
it only remains to give an explicit construction for the  soft functions $S^{(\rm f)}$
in Eq.\ (\ref{eq:facthatsig}), in terms of matrix elements of composite
operators.  We shall proceed in two steps.

We start by representing the coupling of soft gluons to the partons
involved in the hard scattering,
of flavors
$f_A$, $f_B$, $f_1$ and $f_2$,
by ordered exponentials, also known as Wilson lines.
We shall introduce the notation
\beq
\Phi_{\beta}^{(f_i)}({\lambda}_2,{\lambda}_1;x)
=
P\exp\left(-ig\int_{{\lambda}_1}^{{\lambda}_2}d{\eta}\; {\beta}{\cdot} A^{(f_i)} ({\eta}{\beta}+x)\right)\, ,
\label{eq:wilson} 
\eeq
with $A^{(f_i)}$ the gauge field, represented as a matrix
in the representation of flavor $f_i$, $i=A,B,1$ or $2$,
of the gauge group $SU(3)$, and $\beta$ the velocity four-vector
of the parton whose interactions
with soft gluons are being approximated. The operator $P$ orders group 
products in the same sense as the ordering in the integration
variable $\lambda$, with the $A$'s with lower values
of $\lambda$ to the right.  The difference in $\Phi_\beta^{(q)}$ and $\Phi_\beta^{(\bar{q})}$
is in the matrices $A^{(q)}=A^\mu_a(\lambda_a/2)$ and
$A^{(\bar{q})}=A^\mu_a(-\lambda^*_a/2)$, with $\lambda_a$ the Gell-Mann
matrices. Note that this notation differs slightly from that
in Ref.\ \cite{KS}.

As we have observed above, these ordered exponentials
summarize not only the coupling of soft gluons to a single quark or
hard gluon line, but also to an entire jet connected to the
hard scattering by such a parton line \cite{CSSrv,CoSo81,CoSt80}.
By connecting these Wilson lines at a 
local vertex, we construct an eikonal
nonlocal operator, which describes 
the emission of soft radiation, due to both incoming
and outgoing hard partons.
We denote the resulting nonlocal operator as $w_I^{(\a)}$, 
\beqa
w_I^{(\a)}(x)_{\{c_k\}}
&=& 
\sum_{\{d_i\}}
\Phi_{\beta_2}^{(f_2)}(\infty,0;x)_{c_2,d_2}\; 
\Phi_{\beta_1}^{(f_1)}(\infty,0;x)_{c_1,d_1}\cr
&\ &\times
\left( c_I^{(\a)}\right)_{d_2d_1,d_Bd_A}\; 
\Phi_{\beta_A}^{(f_A)}(0,-\infty;x)_{d_A,c_A}
\Phi_{\beta_B}^{(f_B)}(0,-\infty;x)_{d_B,c_B}\, ,
\label{eq:wivertex}
\eeqa
with the $\beta_i$ the four-velocities of the Wilson lines that represent the initial-
and final-state jets.
The color tensor $\left(c_I^{(\a)}\right)_{d_2d_1,d_Bd_A}$ describes 
the couplings of the ordered exponentials
with each other in color space. 
 
We use the operator $w_I$ to define a dimensionless ``eikonal cross section", describing the emission
of gluons by the ordered exponentials, 
\beqa
\sigma_{LI}^{(\a,{\rm eik})}\left(\frac{wM_{JJ}}{\mu},\alpha_s(\mu^2),\epsilon\right)
&=&
\sum_{\xi}\, \delta\left(w-w(\xi,\delta_i)\right)
\nonumber\\
&\ & \times 
\langle0|{\bar T}\left[ \left(w^{(\a)}_L(0)\right){}^{\dagger}_{\{b_i\}}\right]|\xi{\rangle}
{\langle}\xi|T\left[w_I^{(\a)}(0)_{\{b_i\}}\right]|0 \rangle \, ,
\label{eq:eikcs}
\eeqa
where $\xi$ designates a set of intermediate states, 
whose contributions to the weight are given by $w(\xi,\delta_i)$.
Recalling from Eq.\ (\ref{psubisquare})
 that the final-state jet masses are linear in the momenta of particles
emitted into the cones, we may identify the contribution to the
total weight from state $\xi$ as
\beq
w(\xi,\delta_i)={\sqrt{2}(\beta_1\cdot  k'_1+\beta_2 \cdot k'_2)+2k'{}^0\over M_{JJ}}
\eeq
for $M_{JJ}^2=2p_1\cdot p_2$, where $k'_i$ is the momentum
emitted into the $i$th jet cone, while $k'{}^0$ is the energy
emitted outside the cones.  
Similarly, for $M_{JJ}^2=(p_1+p_2)^2$,
we have
\beq
w(\xi,\delta_i)={2k'{}^0 \over M_{JJ}}\, .
\eeq
Defined in this fashion, the eikonal cross section contains 
both collinear and ultraviolet divergences, which 
will be treated by factorization and
renormalization below.  

Setting aside the ultraviolet
divergences for the moment, we note that factorization implies that
this cross section is a good picture of the emission of soft
radiation as a result of the hard scattering, at least 
for soft quanta that are outside the cones of the final-state jets,
and not collinear to the incoming partons.  
Inside the cones, or in the
directions of the incoming partons, the collinear divergences of the eikonal cross section 
are
qualitatively the same as those in the full partonic cross  section,
but quantitatively different in general.  The differences are due to
the original eikonal approximation, necessary to factorize soft emission
from the jets.  

\subsection{The soft function}

To extract the infrared safe soft function $S^{(\a)}_{LI}$
from the eikonal cross section Eq.\ (\ref{eq:eikcs}),
we separate collinear and infrared divergences
in the eikonal cross section.
Contributions from collinear quanta are by construction incorporated
in the functions $\psi_i$ and $J_i$ in Eq.\ (\ref{eq:facthatsig}).
To include these regions in $S^{(\a)}_{LI}$ would be double counting.

We avoid this double counting by eliminating
collinear singularities associated
with the initial-state and final-state jets from $S^{(\a)}_{LI}$.  
The observation that makes this procedure possible is that the eikonal cross section
may be factored into initial-state and final-state jets, and a left-over ``reduced"
soft function, $S$, in the same manner as the full partonic
cross section.  
This, ``eikonal", factorization is illustrated in Fig.\ \ref{eikfactfig}.
The result is a convolution in the weights of final-state partons
associated with each of these functions.  
By analogy to Eqs.\ (\ref{functionws1}) and (\ref{functionws2}), these weights 
are given by
\beqa
w_A'={2k_A'{}^0\over M_{JJ}}\quad &\ & w_B'={2k_B'{}^0\over M_{JJ}}\nonumber\\
w_1'={\sqrt{2}\beta_1\cdot k_1'+2k_1'{}^0\over M_{JJ}}\quad &\ & 
w_2'={\sqrt{2}\beta_2 \cdot k_2'+2k_2'{}^0\over M_{JJ}}\nonumber\\
w_S' &=&{2k_S'{}^0\over M_{JJ}}
\label{eikws1}
\eeqa
for $M_{JJ}^2=2p_1\cdot p_2$, and by
\beqa
w_A'={2k_A'{}^0\over M_{JJ}}\quad &\ & w_B'={2k_B'{}^0\over M_{JJ}}\nonumber\\
w_1'={2k_1'{}^0\over M_{JJ}}\quad &\ & 
w_2'={2k_2'{}^0\over M_{JJ}}\nonumber\\
w_S'&=&{2k_S'{}^0\over M_{JJ}}\, ,
\label{eikws2}
\eeqa
for $M_{JJ}^2=(p_1+p_2)^2$,
where we have used Eq.\ (\ref{psubisquare}) to parameterize the contribution of
the eikonal jet functions to the corresponding jet invariant masses.
The primes simply indicate that these variables refer to a 
convolution for the eikonal, rather than the full, cross section. 

The soft function $S_{LI}^{(\a)}$ found by factoring $\sigma^{(\a,{\rm eik})}$ is
exactly the same soft function found from the factorization of
the full partonic cross section, Eqs.\ (\ref{prelimfact})-(\ref{KpsipsiS}) above,
because the soft radiation is insensitive to the internal
structure of the jets and the hard scattering.
Thus we have, by analogy to Eq.\ (\ref{KpsipsiS}) and (\ref{eq:facthatsig}),
\beqa
\sigma_{LI}^{(\a,{\rm eik})}\left(\frac{wM_{JJ}}{\mu},\Delta y,\alpha_s(\mu^2),\epsilon\right)
&=& 
\nonumber\\
&\ & \hspace{-35mm} 
\int_0^1 dw'_Adw'_Bdw'_1dw'_2dw'_{S}\; \delta\left(w-w'_1-w'_2-w'_A-w'_B-w'_S\right)
\nonumber\\
&\ &\hspace{-30mm}
\times \prod_{c=A,B} {j}^{(f_c)}_{\rm{IN}}\left({w'_cM_{JJ}\over \mu},\alpha_s(\mu^2),\epsilon\right)\;
\, 
\prod_{d=1,2} {j}^{(f_d)}_{\rm{OUT}}
\left({w'_dM_{JJ}\over \mu},\alpha_s(\mu^2),\delta_d\right)
\nonumber\\
&\ & \hspace{-30mm} \times {S}^{(\a)}_{LI}\left({w'_{S}M_{JJ}\over \mu},\Delta y,\alpha_s(\mu^2)\right)\, .
\label{Stojjsprime}
\eeqa
$j_{\rm IN}^{(f_c)}$ and $j_{\rm OUT}^{(f_d)}$ are initial-state and final-state jet
eikonal functions, respectively, which summarize the dynamics of
gluons collinear to the Wilson lines of the eikonal cross
section.  They can be given specific operator definitions. 

Consider first the eikonal distributions for the initial-state jets, $i=A,B$.
The phase space for the initial-state eikonal jets is defined 
by the total energy that they emit into the final state (see Eqs.\ (\ref{eikws1})
and (\ref{eikws2})).  As a result, their definitions
are similar to those for the full center-of-mass
distributions, Eq.\ (\ref{psidef}), 
\beqa
j^{(f_i)}_{\rm{IN}}\left({w'_iM_{JJ}\over \mu},\alpha_s(\mu^2),\epsilon\right) 
&=&
{M_{JJ}\over 2\pi}
\int_{-\infty}^\infty dy_0\ {\rm e}^{-iw'_iM_{JJ}y_0} \nonumber \\
&\ & \hspace{-15mm} \times 
\langle 0|\; {\rm Tr}\bigg\{\; {\bar T}[\Phi^{(f_i)}_{\beta_i}{}^\dagger(0,-\infty;y)]
T[\Phi^{(f_i)}_{\beta_i}(0,-\infty;0)]\; \bigg\}\; |0\rangle\, ,
\label{eq:eikjet}
\eeqa
with $y^\nu=(y_0,\vec{0})$ a vector at the spatial origin. As in Eq.\ (\ref{psidef}),
$\epsilon$ in the arguments of $j^{(f_i)}_{\rm{IN}}$ denotes collinear singularities.

Similarly, the collinear dynamics 
of the eikonal final-state
jets are summarized by matrix elements analogous to Eq.\ (\ref{eq:outjet}),
\beqa
j^{(f_i)}_{{\rm{OUT}}}\left({w'_iM_{JJ}\over \mu},\alpha_s(\mu^2),\delta_i\right) &=&
\sum_{\xi}\; \delta\left(w'_i-w(\xi,\delta_i)\right)
\nonumber\\
&\ & \hspace{-25mm} \times 
\langle 0|\; {\rm Tr}\bigg\{\; {\bar T}[\Phi^{(f_i)}_{\beta_i}{}^\dagger(\infty,0;0)]
|\xi \rangle \langle \xi|
 T[\Phi^{(f_i)}_{\beta_i}(\infty,0;0)]\; \bigg\}\; |0\rangle\, , 
\label{eq:eikoutjet}
\eeqa
with $i=1,2$.  Here again $w(\xi,\delta_i)$ is 
given by either Eq.\ (\ref{eikws1}) or (\ref{eikws2}).

We construct (moments of) the soft function by dividing the moments of 
the eikonal cross section (\ref{Stojjsprime})
by the product of moments of the eikonal jets, Eqs.\ (\ref{eq:eikjet}) and (\ref{eq:eikoutjet}),
\beqa
{\tilde S}^{(\a)}_{LI}\left({M_{JJ}\over N\mu},\Delta y,\alpha_s(\mu^2)\right)
&=&
{{\tilde \sigma_{LI}}^{(\a,{\rm eik})}\left({M_{JJ}\over N\mu},\Delta y,\alpha_s(\mu^2),\epsilon\right)
\over
{\tilde j}^{(f_A)}_{\rm{IN}}\left({M_{JJ}\over N\mu},\alpha_s(\mu^2),\epsilon\right)\; 
{\tilde j}^{(f_B)}_{\rm{IN}}\left({M_{JJ}\over N\mu},\alpha_s(\mu^2),\epsilon\right)}\nonumber\\
&\ & \times
\frac{1}
{{\tilde{j}}^{(f_1)}_{{\rm{OUT}}}\left(\frac{M_{JJ}}{N\mu},\alpha_s(\mu^2),\delta_1\right) \: 
{\tilde{j}}^{(f_2)}_{{\rm{OUT}}}\left(\frac{M_{JJ}}{N\mu},\alpha_s(\mu^2),\delta_2\right)}\, .
\nonumber\\
\label{sigeikDY}
\eeqa
This simply
corresponds to the standard factorization procedure 
in axial gauge \cite{CSSrv}.
At one loop, all diagrams that are two-particle reducible
by cutting incoming or outgoing eikonal lines 
with the same four-velocity $\beta_i$ in the amplitude
and its complex conjugate are eliminated from the soft function,
which is then free of collinear divergences.  

Both the eikonal distributions Eq.\ (\ref{eq:eikjet}) 
and (\ref{eq:eikoutjet}), and 
the soft function Eq.\ (\ref{sigeikDY}), are as yet unrenormalized.
Eikonal lines, and vertices made from their
products, may be renormalized in the usual multipicative manner
\cite{eikrenorm1,BottsSt,SotiSt,GK,KK}.
This renormalization will play an important role in
resummation below.

\section{Resummation}

We are now ready to use the renormalization properties of
the jet and soft functions to organize $N$-dependence
in the factorized cross section in moment space, Eq.\ (\ref{sigNHS}). 

\subsection{Renormalization of the soft function}

The soft function ${{\tilde S}_{LI}}^{(\a)}$, as emphasized above, requires
 renormalization as a composite operator.  This renormalization
is a direct consequence of the factorization we have discussed.
The ultraviolet divergences of ${\tilde S}_{LI}^{(\a)}$
arise when approximations appropriate for soft gluons are extended
to all momenta.  The renormalization scale for the soft function,
then, acts as an effective cutoff, separating soft from hard
gluons.  
Indeed, in the product $H_{IL}S_{LI}$, ultraviolet
divergences induced by the factorization cancel against each other
by construction, since the original diagrams have no UV divergences
beyond those taken into account by the usual renormalization of
the theory \cite{CoSo81,St87,BottsSt,CLS}.

Because $H$ and $S$ occur in a product, they must renormalize
multipicatively, with separate renormalization factors
for the amplitude and the complex conjugate \cite{BottsSt,CLS}
\beqa
H^{(\a)}{}^{(0)}_{IL}&=& \prod_{i=A,B,1,2}Z_i^{-1}\; \left(Z_S^{(\a)}{}^{-1}\right)_{IC}H_{CD}\; 
[(Z_S^{(\a)}{}^\dagger)^{-1}]_{DL}\cr
S^{(\a)}{}^{(0)}_{LI}&=&(Z_S^{(\a)}{}^\dagger)_{LB}S_{BA}Z_{S,AI}^{(\a)}{}\, ,
\label{eq:barereno}
\eeqa
where  $Z_i$ is the renormalization
constant of the {\em{i}}th incoming parton (flavor
$f_A$ \dots $f_2$) connecting to $H^{(\a)}$, and  $Z^{(\a)}_{S,CD}$ is a matrix of
 renormalization constants, describing
the renormalization of the soft function.

From Eq.~(\ref{eq:barereno}), the soft function $S^{(\a)}_{LI}$ satisfies 
the renormalization group equation \cite{BottsSt}
\beq 
\left(\mu\frac{\partial}{\partial\mu}+\beta(g)\frac{\partial}{{\partial}g}\right)S^{(\a)}_{LI}=
-(\Gamma_S^{(\a)}{}^\dagger)_{LB}S^{(\a)}_{BI}-S^{(\a)}_{LA}(\Gamma^{(\a)}_S)_{AI}\, ,
\label{eq:resoft}
\eeq
where we have introduced a soft anomalous dimension matrix, $\Gamma^{(\a)}_S$, which is 
computed directly from the
UV divergences of the soft function. We can compute the matrix in a minimal 
subtraction renormalization scheme,
taking $\epsilon=\epsilon_{UV}=4-D$, with $D$ the number of space-time dimensions.
One-loop anomalous dimensions are then given by
\beq
\Gamma^{(\a)}_S (g)=-\frac{g}{2} \frac {\partial}{{\partial}g}{\rm Res}_{\epsilon 
\rightarrow 0} Z^{(\a)}_S (g, \epsilon)\, .
\label{eq:gendefano}
\eeq
The determination and analysis of $\Gamma^{(\a)}_S$ has been carried out in the case
of heavy quark production in \cite{KS}; its calculation for massless
quarks \cite{BottsSt} and gluons at one loop
will  be the subject of a companion paper.  In any case, the
solution to Eq.\ (\ref{eq:resoft}) takes the form
\beqa
{\rm Tr}\Bigg\{ 
H^{(\a)}\left({M_{JJ}\over\mu},\Delta y,\alpha_s(\mu^2)\right)\;
{\tilde S}^{(\a)} \biggl ({M_{JJ}\over N\mu},\Delta y,\alpha_s(\mu^2) \biggr )\Bigg\}
&\ &\nonumber\\
&\ & \hspace{-70mm}
= {\rm Tr}\Bigg\{ 
H^{(\a)}\left({M_{JJ}\over\mu},\Delta y,\alpha_s(\mu^2)\right)\nonumber\\
&\ & \hspace{-70mm}\ \  \times 
\bar{P} \exp \left[\int_\mu^{M_{JJ}/N} {d\mu' \over \mu'}\; 
\Gamma_S^{(\a)}{}^\dagger\left(\alpha_s(\mu'^2)\right)\right]\nonumber\\
&\ & \hspace{-70mm}\ \  \times
{\tilde S}^{(\a)} \biggl (1,\Delta y,\alpha_s\left(M_{JJ}^2/N^2\right) \biggr )\nonumber\\
&\ & \hspace{-70mm}\ \  \times 
P \exp \left[\int_\mu^{M_{JJ}/N} {d\mu' \over \mu'}\; \Gamma_S^{(\a)}\left(\alpha_s(\mu'^2)\right)\right]
\Bigg\}\, ,
\label{softsoln}
\eeqa
where the trace is taken in the space of color structures.  The symbols $P$ and $\bar{P}$ refer
to path-ordering in the same sense as the integration variable $\mu'$ and against it, respectively
(for example, $P$ orders $\Gamma_S^{(\a)}(\alpha_s(\mu^2))$ to the far right
and $\Gamma_S^{(\a)}(\alpha_s(M_{JJ}^2/N^2))$ to the far left).
As usual, at leading logarithmic accuracy in $\mu$, we can simplify this result to
a sum of exponentials, by choosing a basis in which the matrix $\Gamma_S^{(\a)}$ is diagonal.

\subsection{Initial-state jets}  

Starting from Eq.\ (\ref{sigNHS}), we recall the
resummed expression
for the prefactor $(\psi/\phi)$ in Eq.\ (\ref{sigNHS}),
 the exponentiation of whose $N$-dependence
follows, as  usual, from its factorization
properties \cite{St87,CLS}.  The general expression for moments of the ratios of
the functions $\psi$ and $\phi$, evaluated at the common scale $\mu=M_{JJ}$, is \cite{KS}
\beq
\left[{ {\tilde{\psi}}_{f/f}(N,1,\epsilon)
 \over
 {\tilde{\phi}}_{f/f}(N,M_{JJ}^2,\epsilon)}
 \right] 
= R_{(f)}\left(\alpha_s(M^2_{JJ})\right)\; \exp \left[E_{(f)}(N,M_{JJ})\right]\, ,
\label{EAEB}
\eeq
where
\beqa
E_{(f)}\left(N,M_{JJ}\right)
&=&
-\int^1_0 dz \frac{z^{N-1}-1}{1-z}\; 
\Bigg \{\int^{(1-z)^{m_S}}_{(1-z)^2} \frac{d\lambda}{\lambda} 
A_{(f)}\left[\alpha_s(\lambda M_{JJ}^2)\right]\nonumber\\
&\ &   \hspace{-5mm}
+ B_{(f)}\left[\alpha_s((1-z)^{m_s} M_{JJ}^2)\right]
 +\frac{1}{2}\nu^{(f)}\left[\alpha_s((1-z)^2 M_{JJ}^2)\right]  \bigg \}\, .
\label{omegaexp}
\eeqa
The parameter $m_S$ and  the resummed coefficients $B_{(f)}$ depend on the 
factorization scheme, that is, on the definition of 
$\phi_{f/f}$.  The results are rather different for DIS and
${\rm {\overline MS}}$ schemes, in particular \cite{CT1,CLS}.  This difference 
must be 
compensated for by differences in the 
parton distributions themselves.
With DIS ($\overline{\rm MS}$) factorization schemes,
we have $m_S=1\, (0)$ in Eq.\ (\ref{omegaexp}).
In Eq.\ (\ref{EAEB}),
$R_{(f)}(\alpha_s)$ is an $N$-independent function of the coupling, 
which can be  normalized to unity at zeroth order.

The $A_{(f)},B_{(f)}$ and $\nu^{(f)}$ are finite functions of their arguments. 
To reach next-to-leading order accuracy in $\ln(N)$, we need \cite{CT1,CSt}
\beq
A_{(f)}(\alpha_s) = C_f\left ( {\alpha_s\over \pi} 
+\frac{1}{2} K \left({\alpha_s\over \pi}\right)^2\right )\, ,
\label{g1def}
\eeq
with $C_f=C_F\ (C_A)$ for an incoming quark (gluon), and 
with $K$ given by \cite{KT}
\beq
K= C_A\; \left ( {67\over 18}-{\pi^2\over 6 }\right ) - {5\over 9}n_f\, ,
\label{Kdef}
\eeq
where $n_f$ is the number of quark flavors.  
$B_{(f)}$ is given for quarks in the DIS scheme by
\beq
B_{(q)}(\alpha_s)=-{3\over 4}C_F\; {\alpha_s\over\pi}\, , 
\eeq
while it vanishes in the $\overline {\rm MS}$ scheme for quarks and gluons.
(The DIS scheme is normally only applied to quarks,
although extended definitions for gluons are possible
\cite{OwensTung}.)  
Finally, the lowest-order approximation to $\nu^{(f)}$, which is scheme-independent, is \cite{KS}
\beq
\nu^{(f)}=2C_f\; {\alpha_s\over\pi}\, .
\eeq

The results just discussed are useful, but not quite adequate for
our needs, because they assume that $\mu=M_{JJ}$, as would
be appropriate for Drell-Yan cross sections.  For jet cross
sections, we generally need additional freedom to choose the
factorization scale.  
To change the scale $\mu$, we need the
renormalization group behavior of the parton distributions
$\psi$ and $\phi$, whose ratio enters Eq.\ (\ref{sigNHS}).

The center of mass distribution $\psi$ requires no overall
renormalization as a composite operator \cite{St87}.     
From its definition, Eq.\ (\ref{psidef}),
we see that $\psi$, and each of its moments, renormalizes multiplicatively,
because it is the matrix element of a product of renormalized
operators.  As a result, it obeys the renormalization group equation
\beq
\mu{d\tilde{\psi}_{f/f}(N,M_{JJ}/\mu,\epsilon) \over d \mu}
=2\gamma_f(\alpha_s(\mu^2))\; \tilde{\psi}_{f/f}(N,M_{JJ}/\mu,\epsilon)\, ,
\label{psirg}
\eeq
with $\gamma_f$ the anomalous dimension of the field of flavor $f$,
which is, of course, independent of $N$.  

The dependence of the light-cone distribution $\tilde{\phi}_{f/f}$ on 
the factorization scale $\mu$ is slightly more complex than 
for $\psi$, and depends on the factorization scheme that we choose.
The simplest case is the $\overline{\rm MS}$ scheme.  Each of the moments of
the $\overline{\rm MS}$ version of $\phi$ obeys a renormalization
group equation with the anomalous dimension of the color-diagonal
splitting function for that flavor,
\beq
\mu{d\tilde{\phi}_{f/f}(N,\mu^2,\epsilon) \over d \mu}
=2\gamma_{ff}(N,\alpha_s(\mu^2))\; \tilde{\phi}_{f/f}(N,\mu^2,\epsilon)\, .
\label{phirg}
\eeq
For the $\overline{\rm MS}$ distribution, this relation holds by
definition, because the distributions are defined as the
matrix elements of operators on the light-cone, whose renormalization
is described by the splitting functions \cite{CoSoPDF,QQbarCMNT,CSt}.
Only the flavor-diagonal evolution survives in the
large-$N$ limit, because only
color diagonal splitting functions are singular for $x\rightarrow 1$.    

The factorization scale dependence now may be controlled, by
the two evolution equations (\ref{psirg}) and (\ref{phirg}).
The $\overline{\rm MS}$
scheme expression that includes the $\mu$-dependent prefactor in Eq.\ (\ref{sigNHS}),
thus generalizing Eq.\ (\ref{EAEB}), is
\beqa
\left[{ {\tilde{\psi}}_{f/f}(N,M_{JJ}/\mu,\epsilon)
 \over
 {\tilde{\phi}}_{f/f}(N,\mu^2,\epsilon)}
 \right]_{\overline{\rm MS}}
&=& R_{(f)}\left(\alpha_s(\mu^2)\right)\; \exp{\left[E_{(f)}(N,M_{JJ})\right]}\nonumber\\
&\ & \times
\exp \left \{ -2\int_\mu^{M_{JJ}}{d\mu'\over\mu'}\; \left [\gamma_f(\alpha_s(\mu'{}^2))
-\gamma_{ff}(N,\alpha_s(\mu'{}^2)) \right]\right\}\, .
\nonumber\\
\label{psiphiscale}
\eeqa
This expression enables us to change factorization scales in the
resummed cross section.  The corresponding factor for DIS scheme distributions
is easily found by comparing the ratio 
in Eq.\ (\ref{EAEB}) in DIS and $\overline{\rm MS}$
schemes.   The effect of the DIS scheme is to produce
an extra factor in Eq.\ (\ref{psiphiscale}), 
\beqa
\left[{ {\tilde{\psi}}_{f/f}(N,M_{JJ}/\mu,\epsilon)
 \over
 {\tilde{\phi}}_{f/f}(N,\mu^2,\epsilon)}
 \right]_{{\rm DIS}}
&=& R_{(f)}\left(\alpha_s(\mu^2)\right)\; \exp{\left[E_{(f)}(N,M_{JJ})\right]}\nonumber\\
&\ & \hspace{-7mm} \times
\exp \left \{ -2\int_\mu^{M_{JJ}}{d\mu'\over\mu'}\; \left [\gamma_f(\alpha_s(\mu'{}^2))
-\gamma_{ff}(N,\alpha_s(\mu'{}^2)) \right]\right\}
\nonumber\\
&\ & \hspace{-7mm} \times \exp \Bigg [ \int^1_0 dz \frac{z^{N-1}-1}{1-z}\; 
\bigg \{\int^1_{(1-z)} \frac{d\lambda}{\lambda} 
A_{(f)}\left[\alpha_s(\lambda \mu^2)\right]\nonumber\\
&\ &  \quad \quad \quad \quad 
- B_{(f)}\left[\alpha_s((1-z)\mu^2) \right] \bigg \}   \Bigg]\, .
\nonumber\\
\label{psiphiscaleDIS}
\eeqa
This difference may also be formulated in terms of a slightly modified evolution
equation for DIS distributions, which takes into account 
their logarithmic moment dependence \cite{GSDIS}. 

\subsection{Final-state jets}

The remaining factors in Eq.\ (\ref{sigNHS}) are the final-state
jet functions $J^{(f_i)}$.  The
resummation of their $N$-dependence
depends critically on the definition of the
dijet cross sections.   For illustrative purposes, we have identified
the choices $M_{JJ}=(p_1+p_2)^2$ and $2p_1\cdot p_2$.  We begin with
the latter.

{\it The case $M_{JJ}=2p_1\cdot p_2$}.  
In this case, the jet function contributes to the
overall weight through the mass of the cone-jet, $p_i^2$,
and also through the total energy of particles emitted outside
the cone.  Thus, recalling Eq.\ (\ref{functionws1}), we may
write $J^{(f_i)}$ as an integral over these variables
at fixed $w_i$,
\beqa
J^{(f_i)}\left(w_i,{M_{JJ}\over \mu},\alpha_s(\mu^2),\delta_i\right)
&=&
\int dp_i^2dk_i^0\, 
\delta\left(w_i-{p_i^2+2M_{JJ}k_i^0\over M_{JJ}^2}\right)
\nonumber\\
&\ & \quad \times I^{(f_i)}\left({M_{JJ}\over \mu},{p_i^2\over\mu^2},{k_i^0\over\mu},\delta_i\right)\, ,
\label{JtoI}
\eeqa
with $I^{(f_i)}$ a density in the invariant mass inside the cone, and energy outside.
This makes its analysis a bit more complicated
than the cases studied in Ref.\ \cite{CLS}.  

In the limit of very small $p_i^2$, however, we may 
factorize soft gluons emitted at angles much larger
than $\sqrt{p_i^2/M_{JJ}^2}$ from a function 
that describes the dynamics of collinear partons.  This factorization can be carried
out by the procedure of Ref.\ \cite{CoSo81}, referred to in Sec.\ 3.2 above.
As discussed above, couplings of soft gluons to the jet are first
replaced by an eikonal, or ``soft" approximation.  Suppose the
jet velocity is $\beta_i$.
In the soft approximation,  soft gluon momenta carried by jet lines 
are approximated by their 
``opposite-moving" component $\beta_i\cdot q$, and similarly
for soft gluon polarizations at the vertices where they
are emitted by jet lines. 
  
After
the use of Ward identities, soft gluons factorize in the soft 
approximation \cite{CSSrv,CoSo81},
and we derive a convolution of the form,
\beqa
I^{(f_i)}\left({M_{JJ}\over \mu},{p_i^2\over \mu^2},{k_i^0\over\mu},\delta_i\right)
&=&
\int d(\beta_i\cdot q)\, \bar{\Sigma}^{(f_i)}
\left({M_{JJ}\over \mu},{p_i^2-\sqrt{2}M_{JJ}(\beta_i\cdot q) \over \mu^2}\right)
\nonumber\\
&\ & \hspace{10mm} \times
j_{\rm OUT}^{(f_i)}\left({(\beta_i\cdot q)\over\mu},{k_i^0\over\mu}\right)\, .
\label{jetfact}
\eeqa
Corrections to this relation are free of infrared divergences in the
limit of small $k_i^0$, $(\beta_i\cdot q)$, and hence do not contribute to the leading power 
for $w_i\rightarrow 0$.
The function $j_{\rm OUT}^{(f_i)}$, which absorbs the soft gluon
dynamics is the same eikonal jet
function defined in Eq.\ (\ref{eq:eikoutjet}) and shown in Fig.\ \ref{eikfactfig}.
The function $\bar{\Sigma}^{(f_i)}$ remains dependent on the jet's longitudinal
momentum, and is linked to 
$j_{\rm OUT}^{(f_i)}$ by a convolution in the
small component ($\beta_i\cdot q$) of momentum emitted by that function into the cone.  
In the limit of small $p_i^2$, gluons emitted outside the cone
are all associated with the eikonal function.   The most important feature of
$\bar{\Sigma}^{(f_i)}$ for us is that it is independent of the cone size,
up to corrections that vanish with $p_i^2$.

The convolution in Eq.\ (\ref{jetfact}) simplifies under moments of $J^{(f_i)}$,
Eq.\ (\ref{JtoI}),
with respect to  $w_i$,
\beqa
\int dw_i\; {\rm e}^{-Nw_i}\
J^{(f_i)}\left(w_i,{M_{JJ}\over \mu},\alpha_s(\mu^2),\delta_i)\right)&\ &
\nonumber \\
&\ &  \hspace{-30mm} = \int d(\beta_i\cdot q)\; dk_i^0\; 
{\rm e}^{-N\left({\sqrt{2}(\beta_i\cdot q)+ 2k_i^0\over M_{JJ}}   \right)}
j_{\rm OUT}^{(f_i)}\left({(\beta_i\cdot q)\over\mu},{k_i^0\over\mu}\right)
\nonumber\\
&\ & \hspace{-30mm} \ \ \times\;
\int dp'{}^2\; {\rm e}^{-N\left({p'{}^2\over M_{JJ}^2}\right)}
\bar{\Sigma}^{(f_i)}\left({M_{JJ}\over \mu},{p'{}^2 \over \mu^2}\right)
\nonumber \\
&\ & \hspace{-30mm} =
\tilde{j}_{\rm OUT}^{(f_i)}\left({M_{JJ}\over N\mu}\right)\;
\tilde{\Sigma}^{(f_i)}\left({M_{JJ}\over \mu},{M_{JJ}^2 \over N\mu^2}\right)\, ,
\label{jetmoments}
\eeqa
where in the second line,  $p'{}^2=p_i^2-\sqrt{2}M_{JJ}(\beta_i\cdot q)$.
The moment integrals here and below all have lower limits at zero.  
Dependence on their upper limits is exponentially suppressed in $N$,
and may be neglected. Our task now is to derive the $N$-dependence of the product
$\tilde{j}_{\rm OUT}^{(f_i)}\tilde{\Sigma}^{(f_i)}$.

To find the requisite $N$-dependence, we begin by observing that a very
similar procedure may be applied to the {\it full} axial gauge two-point function,
which we denote as $\Sigma_2^{(f_i)}$.
Again, soft gluon emission away from the jet axis 
factorizes into a convolution form,
\beq
\Sigma_2^{(f_i)}\left({M_{JJ}\over \mu},{p_i^2\over \mu^2}\right)
=
\int d(\beta_i\cdot q)\; 
\bar{\Sigma}^{(f_i)}\left({M_{JJ}\over \mu},{p_i^2-\sqrt{2}M_{JJ}(\beta_i\cdot q) \over \mu^2}\right)\;
j_-^{(f_i)}\left({(\beta_i\cdot q)\over\mu}\right)\, ,
\label{sig2convol}
\eeq
with $j_-^{(f_i)}$ the two-point eikonal function, with its eikonal
line again in the $\beta_i$-direction, evaluated at fixed values of the
$\beta_i\cdot q$ momentum component of emitted partons.   Here $\bar{\Sigma}^{(f_i)}$ is
the {\it same} function as in Eq.\ (\ref{jetfact}) for small $p_i^2$, because
in this limit collinear gluons are emitted only at the 
center of the cone, far from its boundary.
Just as for Eq.\ (\ref{jetmoments}), 
$\Sigma_2^{(f_i)}$ factorizes into a product under moments, this time with
respect to $p_i^2$,
\beqa
\int dp_i^2\; {\rm e}^{-N\left({p_i^2\over M_{JJ}^2}\right)}\
\Sigma_2^{(f_i)}\left({M_{JJ}\over \mu},{p_i^2\over \mu^2}\right)
&=&  
\int d(\beta_i\cdot q)\; {\rm e}^{-N\left({\sqrt{2}(\beta_i\cdot q)\over M_{JJ}}   \right)}
j_-^{(f_i)}\left({(\beta_i\cdot q)\over\mu}\right)
\nonumber\\
&\ & \vspace{-20mm} \times\;
\int dp'{}^2\; {\rm e}^{-N\left({p'{}^2\over M_{JJ}^2}\right)}
\bar{\Sigma}^{(f_i)}\left({M_{JJ}\over \mu},{p'{}^2 \over \mu^2}\right)
\nonumber \\
&=& \vspace{-20mm}
\tilde{j}_{-}^{(f_i)}\left({M_{JJ}\over N\mu}\right)\;
\tilde{\Sigma}^{(f_i)}\left({M_{JJ}\over \mu},{M_{JJ}^2 \over N\mu^2}\right)\, .
\label{2pointmoments}
\eeqa
Comparing Eqs.\ (\ref{jetmoments}) and (\ref{2pointmoments}), we see
that the moments of the cone-jet function $J^{(f_i)}$ 
with respect to $w_i$
are closely related to the moments of the full two-point function
$\Sigma_2$ with respect to $p_i^2$,
\beq
\int dw_i\; {\rm e}^{-Nw_i}\
J^{(f_i)}(w_i)
=
\left[
{\tilde{j}_{\rm OUT}^{(f_i)}\left({M_{JJ}\over N\mu}\right)
\over
 \tilde{j}_-^{(f_i)}\left({M_{JJ}\over N\mu}\right) }\right]
\, 
\int dp_i^2\; {\rm e}^{-N\left({p_i^2\over M_{JJ}^2}\right)}\;
\Sigma_2^{(f_i)}(p_i^2)\, .
\label{joutmoment}
\eeq
The $N$-dependence of the moments of the jet function are thus
given by the moments of the full two-point function, 
times the ratio
of moments of the two eikonal functions $j_{\rm OUT}$ and $j_-$.
We readily find their $N$-dependence as follows.

Moments of a cut two-point function with respect to its
invariant mass were resummed explicitly in
Ref.\ \cite{St87}.  The $\ln (N)$ dependence
exponentiates in analogy to Eqs.\ (\ref{EAEB})
and (\ref{omegaexp}).  Normalizing the two-point function
to $\delta(p_i^2/M_{JJ}^2)$ at lowest order, we have
\beq
\int dp_i^2\; {\rm e}^{-N\left({p_i^2\over M_{JJ}^2}\right)}\
\Sigma_2^{(f_i)}\left({M_{JJ}\over \mu},{p_i^2\over \mu^2}\right) 
= \exp \left[E'_{(f_i)}(N,M_{JJ})\right]\, ,
\label{Eout}
\eeq
where
\beqa
E'_{(f)}\left(N,M_{JJ}\right)
&=&
\int^1_0 dz \frac{z^{N-1}-1}{1-z}\; 
\Bigg \{\int^{(1-z)}_{(1-z)^2} \frac{d\lambda}{\lambda} 
A_{(f)}\left[\alpha_s(\lambda M_{JJ}^2)\right]\nonumber\\
&\ &  \quad \quad 
+ B'_{(f)}\left[\alpha_s((1-z) M_{JJ}^2) \right] \bigg \}\, .
\label{outexp}
\eeqa
The prime on $E'$ indicates the exponent for a final-state jet, while the function
$A_{(f)}$ is the same as in Eq.\ (\ref{g1def}).
The lowest order term in $B'_{(f)}$
may be read off from the one-loop jet function.  
The results include a gauge dependence, which cancels
against a corresponding dependence in the soft anomalous
dimension matrix \cite{BottsSt}.

The other factor in Eq.\ (\ref{joutmoment}),
$\tilde{j}_{\rm OUT}(N)/\tilde{j}_-(N)$
actually affects the moments of the final-state
jet only beyond next-to-leading logarithm in $N$.  This may
be seen as follows.  First, being the sum of
fully eikonal diagrams, both $\tilde{j}_{\rm OUT}$ and
$\tilde{j}_-$ exponentiate in moment space,
according to the general arguments of Ref.\ \cite{Gatheral}.
At the same time, double logarithmic contributions,
which are associated with gluons collinear to
the eikonal lines, match in $\tilde{j}_{\rm OUT}$ and
$\tilde{j}_-$, because the phase space for 
collinear gluons is the same in both functions.
The ratio may thus include at most next-to-leading
logarithms, associated with soft gluons emitted
outside the cone.  
The contribution of next-to-leading logs
in the exponential may then be
determined from a one-loop calculation. 
As for the soft function above, single logarithms are determined
by the renormalization of one-loop virtual  diagrams, which are
identical for $j_{\rm OUT}$ and $j_-$.
An explicit calculation therefore gives no
logarithms at all in the ratio.  

In summary, up next-to-leading logarithm, $j_{\rm OUT}/j_-=1$,
and the moments of the final-state jet in Eq.\ (\ref{joutmoment}) with
$M_{JJ}^2=2p_1\cdot p_2$ are given by
\beq
\tilde{J}^{(f_i)}\left(N,{M_{JJ}\over \mu},\alpha_s(\mu^2),\delta_i\right)
=
\int dw_i\; {\rm e}^{-Nw_i}\
J^{(f_i)}(w_i)
=
\exp \left[E'_{(f_i)}(N,M_{JJ})\right]\, ,
\label{outjetmoments}
\eeq
with $E'_{(f)}$ defined as in Eq.\ (\ref{outexp}), up to
next-to-next-to-leading corrections in the function
$B'_{(f)}$.

An important feature of the exponent
associated with the final-state jet is that
it has the opposite overall sign from the exponent of
Eq.\ (\ref{omegaexp}), associated with the initial
state.  The  leading logarithmic contributions
of the final-state jets are negative, but
are otherwise determined by the same functions
$A_{(f)}(\alpha_s)$.  
Just as the leading-log initial-state
contributions always act to enhance the cross
section, those associated with the
final state always suppress it, by an amount that depends on the partonic
subprocess.
Such singularities correspond to Sudakov suppression
of scattering in the elastic limit \cite{CLS,BottsSt}.  
They are already present in explicit next-to-leading
order calculations of single-jet inclusive cross sections \cite{jetcalc}.
It may be worth pointing out
that for the specific choice of subprocess $q\bar{q}\rightarrow gg$,
the net coefficient of the leading logarithm is
negative in both DIS and $\overline{\rm MS}$
schemes (because $C_A>C_F/2$), in  sharp contrast to
the classic Drell-Yan calculation \cite{St87,CT1}.  

{\it The case $M_{JJ}^2=(p_1+p_2)^2$.}  When we choose $M_{JJ}$
as the total invariant mass of the two-jet system, the 
phase space at partonic threshold is, 
as mentioned in Sec.\ 3.1 above,
changed significantly compared to the previous choice.   In the present case,
the masses of the dijets are not 
forced to zero at threshold (see Eq.\ (\ref{k0to1minusz})).  In fact, the phase space
at partonic threshold consists of all states in which
each of the two jets carries total energy $M_{JJ}/2$, with
equal and opposite momenta.  By dimensional
considerations, the jet masses may increase to the 
order of $\delta^2M_{JJ}^2$. As a result of this larger
phase space, collinear logarithms of $N$ cancel, leaving
only infrared enhancements associated
with soft emission outside the cone, which
appear as single logarithms of $N$ per loop.
The remnants of the dilogarithmic structure 
near threshold are terms of the 
form $\alpha_s^n\; (\ln\delta_i\; \ln N)^n$,
which replace the leading $\alpha_s^n\; (\ln N)^{2n}$
terms of the previous case.

Although the jet functions with $M_{JJ}^2=(p_1+p_2)^2$
involve fewer logs
per loop than for $M_{JJ}^2=2p_1\cdot p_2$, the
situation is in some sense more complicated 
than before.  This is 
because of configurations in which the jets, although
of minimum energy, are nevertheless massive.
In these regions of phase space, soft emissions near the edge of the
cone may resolve the flow of color within the
cone  \cite{CoCoh}, carried, for  example, by fast partons
separated by angles of order $\delta_i$.
The couplings of soft gluons to such
configurations may not be approximated by their
coupling to a single eikonal line at the 
center of the cone.  This approximation was accurate to all logs 
for $M_{JJ}^2=2p_1\cdot p_2$, because in that case the limit $N\rightarrow\infty$
forces the jet mass to zero, and hence the 
energetic partons to the center of the cone.
Nevertheless, the approximation in which 
soft gluons are emitted by a single eikonal line at
the center of the jet still captures the next-to-leading,
$\alpha_s^n\; \ln^nN$, contributions.
Any such eikonal cross section, with phase space that is symmetric
in all particles, exponentiates \cite{Gatheral,epemjetexp},  
\beqa
\tilde{J}^{(f)}\left(N,{M_{JJ}\over \mu},\alpha_s(\mu^2),\delta_i\right)
&=& \exp[E'_{(f)}(N,M_{JJ})]\nonumber\\
E'_{(f)}(N,M_{JJ}) &=&
\exp \left[\int_\mu^{M_{JJ}/N} {d\mu' \over \mu'}\; \; 
C'_{(f)}(\alpha_s(\mu'{}^2)) \right]\, ,
\label{Cexp}
\eeqa
where $C'_{(f)}(\alpha_s)$ is a  finite perturbative
series whose first term may be read off from
a one-loop calculation.  Beyond these next-to-leading logarithms,
however, corrections to Eq.\ (\ref{Cexp}) give
a rather complicated sum of terms, each representing
the coherent radiation due to
a set of eikonal lines along arbitrary directions
within the cone.  

Again to avoid
a proliferation of symbols, we use the
same notation for the exponent here as for 
$M_{JJ}^2=2p_1\cdot p_2$, although the leading behavior
in the two cases is quite different.   
For $M_{JJ}^2=(p_1+p_2)^2$, the leading logarithmic
behavior in moments (and therefore the leading
distributions in momentum space) are not affected
by the final-state jets, and thus retain the
same (positive) contributions encountered in
Drell-Yan cross sections.

We are now finally ready to assemble all the 
pieces necessary to write down a resummed 
cross section for dijet production.  

\subsection{The resummed cone-dijet cross sections}

The characteristically nonabelian 
aspect of resummation in our case lies in the evolution of the soft functions.
Recall that because the anomalous dimension matrix is not diagonal in general,
solutions to the evolution equation (\ref{eq:resoft}) are ordered, rather than simple,
exponentials.  Substituting the solution (\ref{softsoln})
to Eq.\ (\ref{eq:resoft})
into the cross section in moment space, Eq.\ (\ref{sigNHS}), and
using Eqs.\ (\ref{EAEB}),
(\ref{psiphiscale}) and (\ref{outjetmoments}) (or Eq.\ (\ref{Cexp})),
we find an expression that organizes all logarithms of $N$ at leading power in $N$,
in $\overline{\rm MS}$ factorization scheme,
\beqa
\tilde{{\sigma}}_{\a}(N) &=& R_{(f)}\; 
\exp \Bigg \{ \sum_{i=A,B} \bigg[ E_{(f_i)}(N,M_{JJ}) 
\nonumber\\
&\ & \hspace{20mm} 
-2\int_\mu^{M_{JJ}}{d\mu'\over\mu'}\; 
\big [\gamma_{f_i}(\alpha_s(\mu'{}^2))-\gamma_{f_if_i}(N,\alpha_s(\mu'{}^2)) \big] \bigg] \Bigg\}
\nonumber \\
&\ &\times\; \exp \Bigg \{  \sum_{j=1,2}E'_{(f_j)}(N,M_{JJ}) \Bigg\}\nonumber\\
&\ &\times\; {\rm Tr}\Bigg\{ 
H^{(\a)}\left({M_{JJ}\over\mu},\Delta y,\alpha_s(\mu^2)\right)\nonumber\\
&\ & \times\;
\bar{P} \exp \left[\int_\mu^{M_{JJ}/N} {d\mu' \over \mu'}\; \Gamma_S^{(\a)}{}^\dagger\left(\alpha_s(\mu'^2)\right)\right]\;
{\tilde S}^{(\a)} \biggl (1,\Delta y,\alpha_s\left(M_{JJ}^2/N^2\right) \biggr )\nonumber\\
&\ & \times\; 
P \exp \left[\int_\mu^{M_{JJ}/N} {d\mu' \over \mu'}\; \Gamma_S^{(\a)}\left(\alpha_s(\mu'^2)\right)\right]
\Bigg\}\, ,
\label{sigNHSfinal}
\eeqa
where as above the trace is taken in the space of color structures.  
For DIS scheme, the only change is the additional factor given 
in Eq.\ (\ref{psiphiscaleDIS}).

Without going into detail at this point, certain general features of 
the resummed cross section may be readily identified.  The first
exponential factors, $\exp[E_{(f_i)}(N)]$, serve, as in Drell-Yan cross
sections, to enhance the cross section.  
The next factors, involving
the anomalous dimensions of the parton fields and splitting functions,
govern factorization scale dependence.   
The third factor, associated with final-state jets, always
acts to suppress the cross section.  We have noted already
how the choice of jet algorithm can influence their size.
The hard-scattering functions
remain dependent on the scattering angle.
 The most
interesting factors, however, are the ordered exponentials of the
soft anomalous dimension matrix.  They
depend on the scattering angle and they distinguish the roles of different color structures
in the hard scattering
through coherent gluon emission,  linking initial-
and final-state partons.  

The logarithmic
accuracy of the resummation depends upon the  order to which
the hard scatterings, and the anomalous dimensions, have been calculated.
In a paper to follow this one, we shall compute the
leading-order anomalous dimension matrices for each flavor
combination in QCD.  This will enable us
to present explicit resummed dijet cross sections
to next-to-leading order in $\ln (N)$ \cite{KOSip}.

\subsection{Other jet algorithms and kinematics}

Eq.\ (\ref{sigNHSfinal}) is typical of resummed jet cross sections, but its
details depend on our use of cones to define the jets.  Other choices, of course,
are possible, and in cases, preferable.  One of the disadvantages of the
cone criterion is that it requires an extra parameter $\delta_i$ for each jet.
In addition, single-jet, rather than dijet cross sections are often more
convenient experimentally.  Single-jet inclusive cross sections
have slightly different kinematic properties near threshold, compared to dijet,
and therefore result in slightly different resummed expressions \cite{KS}.  The 
extension of threshold resummation to cross sections with 
single-particle kinematics will be discussed in Ref.\ \cite{LOSip}.
Nevertheless, it is clear that the result (\ref{sigNHSfinal})
is much more general than the cone algorithm that we have used.
In fact, all that is necessary to derive such a resummed expression
is a convolution form like Eq.\ (\ref{eq:facthatsig}), in which the weight is linear
in the momenta of radiated gluons. Threshold logarithms in any such weight will be
controlled by the same matrix of anomalous dimensions $\Gamma_S$
that we have identified above.  Differences between different algorithms
will, in general, show up only in the functions $B'_{(f)}$ for the
final-state jets in Eq.\ (\ref{outexp}).

\section{Conclusion}

In this paper, we have shown how to resum threshold logarithms in cone-based dijet
cross sections.  The method may be extended to other jet algorithms, and
the explicit form of the resummed cross section will depend upon the details
of the cross section chosen.  In every case, however, the resummation depends
upon the color structure of the hard scattering.  
As for the Drell-Yan cross section,  the resummation of threshold singularities
depends on the factorization scheme, although the color-dependent contribution
does not.  Our explicit result for the $\overline{\rm MS}$ factorization
scheme is given in Eq.\ (\ref{sigNHSfinal}) above.  It is, in principle, valid
to all logarithmic accuracy, at leading power of $N$ in moment space for
the dijet cross section at fixed $p_1\cdot p_2$, and to next-to-leading
logarithm in $N$ for fixed $(p_1+p_2)^2$.
Its inverse transform to momentum space therefore summarizes singular
distributions in $1-z$ to all logarithmic order in the former case
and to next-to-leading order in the latter.  As for heavy
quark production, the general resummation can only be given in terms 
of ordered exponentials, due to mixing of color exchange tensors by soft gluon
emission.
In a forthcoming paper \cite{KOSip}, we will derive the soft anomalous dimension
matrix to one loop for the full range of flavor scatterings that give
rise to jet production, and give explicit expressions for dijet
cross sections to leading order in the coupling, and to leading and
next-to-leading logarithm in the moment variable. 

\subsection*{Acknowledgements}

We wish to thank Eric Laenen for many helpful conversations.
This work was supported in part by the National Science Foundation,
under grant PHY9722101 and by the PPARC under grant GR/K54601.

\newpage

\newpage

\newpage

\begin{figure}
\centerline{\epsffile{fig3.epsi}} 
\caption{Factorized form of the cross section
into initial-state and final-state jet functions ($\psi$ and $J$, respectively)
and the color-dependent soft function $S_{LI}^{(\a)}$.  At lowest
order, $S_{LI}^{({\rm f})}$ is given simply by the set of all diagrams in
which a single gluon connects any two ordered eikonal
lines moving in different directions.  The construction
of $S_{LI}^{({\rm f})}$ beyond lowest order in discussed in Sec.\ 3.2.}
\label{factfig}
\end{figure}

\newpage

\begin{figure}
\centerline{\epsffile{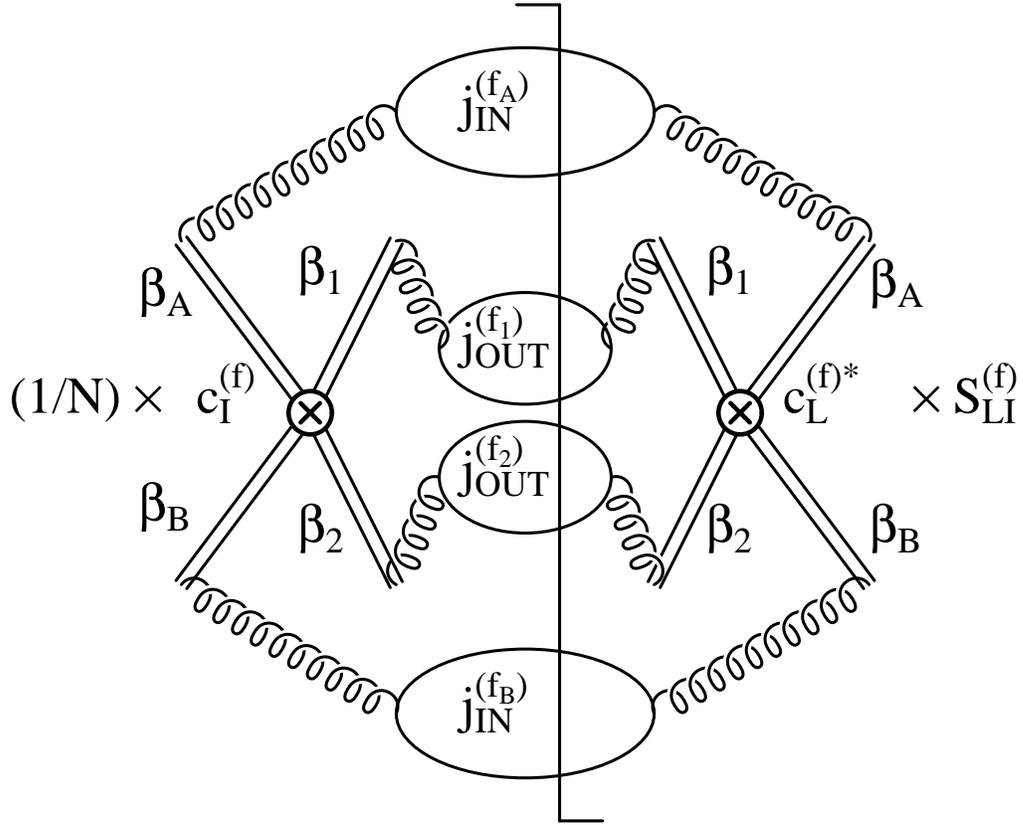}} 
\caption{Factorized eikonal cross section. With an appropriate
choice of color normalization factor $1/{\cal N}$, the soft function
$S^{({\rm f})}_{LI}$ is the same as in Fig.\ {\protect \ref{factfig}}.  The eikonal jet functions
$j_{\rm OUT}$ and $j_{\rm IN}$ are defined in Eqs.\ ({\protect \ref{eq:eikjet}})
and ({\protect \ref{eq:eikoutjet}}), respectively.}
\label{eikfactfig}
\end{figure}

\end{document}